\documentclass{nature}
 
\usepackage[version=3]{mhchem} 
\usepackage[T1]{fontenc}       
\usepackage{color}
\usepackage{textcomp}
\usepackage{amsmath}
\usepackage{amssymb}
\usepackage[draft]{changes}
\definechangesauthor[color=orange]{DW}
\usepackage{graphicx}
\usepackage{amsmath}
\usepackage{amssymb}
\usepackage{amsfonts}
\usepackage{textcomp}
\usepackage{bm}
\usepackage{multirow}
\usepackage{float}
\usepackage{placeins}

\usepackage{color}

\usepackage{cancel}

\usepackage[font=footnotesize,labelfont=bf]{caption}
\usepackage{geometry}
 \usepackage[sort, numbers]{natbib}
 \setcitestyle{square}

\usepackage{graphicx}
\makeatletter
\let\saved@includegraphics\includegraphics
\AtBeginDocument{\let\includegraphics\saved@includegraphics}
\renewenvironment*{figure}{\@float{figure}}{\end@float}
\makeatother

\title{Destructive photon echo formation in six-wave mixing signals of a MoSe$_{\boldsymbol{2}}$ monolayer}

\author{\small
Thilo Hahn,$^{1,2}$ Diana Vaclavkova,$^{3}$ Miroslav Bartos,$^{3,4}$ Karol Nogajewski,$^{5}$\\ Marek Potemski,$^{3,5}$ Kenji Watanabe,$^{6}$ Takashi Taniguchi,$^{7}$\\ Pawe\l{} Machnikowski,$^{2}$ Tilmann Kuhn,$^{1}$ Jacek Kasprzak,$^{8}$ Daniel Wigger$^{2,\ast}$}

\begin{document}

\maketitle

\begin{affiliations}\small
 \item Institut of Solid State Theory, University of M\"unster,\\ 48149 M\"unster, Germany
 \item Department of Theoretical Physics, Wroc\l{}aw University of Science and Technology,\\ 50-370~Wroc\l{}aw, Poland
 \item Laboratoire National des Champs Magn\'{e}tiques Intenses, CNRS-UGA-UPS-INSA-EMFL,\\ 38042 Grenoble, France
 \item Central European Institute of Technology, Brno University of Technology,\\ 61200 Brno, Czech Republic
 \item Institute of Experimental Physics, Faculty of Physics, University of Warsaw,\\ 02-093 Warszawa, Poland
 \item Research Center for Functional Materials, National Institute for Materials Science,\\ Tsukuba 305-0044, Japan
 \item International Center for Materials Nanoarchitectonics, National Institute for Materials Science,\\ Tsukuba 305-0044, Japan
 \item Universit\'{e} Grenoble Alpes, CNRS, Grenoble INP, Institut N\'{e}el,\\ 38000 Grenoble, France
 \item[$^\ast$] daniel.wigger@pwr.edu.pl
\end{affiliations}

\begin{abstract}\small

Monolayers of transition metal dichalcogenides display a strong excitonic optical response.  Additionally encapsulating the monolayer with hexagonal boron nitride allows to reach the limit of a purely homogeneously broadened exciton system. On such a MoSe$_{2}$-based system we perform ultrafast six-wave mixing spectroscopy and find a novel destructive photon echo effect. This process manifests as a characteristic depression of the nonlinear signal dynamics when scanning the delay between the applied laser pulses. By theoretically describing the process within a local field model we reach an excellent agreement with the experiment. We develop an effective Bloch vector representation and thereby demonstrate that the destructive photon echo stems from a destructive interference of successive repetitions of the heterodyning experiment.

\end{abstract}

\thispagestyle{empty}

\twocolumn
\footnotesize

\section{Introduction}
The spin echo~\cite{hahn1950spin} is an essential effect in nuclear magnetic resonance spectroscopy and the basis for all sorts of complex radio pulse sequences~\cite{bernstein2004} that are routinely applied in medicine~\cite{hendee1999physics}, chemistry~\cite{hore2015nuclear}, or physics~\cite{zutic2004spintronics, weichselbaumer2020echo, debnath2020self}. While spin resonances are driven by radio frequencies, optical frequencies are required to resonantly excite typical charge transitions; in this regime the analogous phenomenon is called photon echo~\cite{kurnit1964obse}. First demonstrations were performed on ruby crystals~\cite{kurnit1964obse, abella1966phot}, but the photon echo spectroscopy also has a long standing history in semiconductor optics. It has been used to study different types of exciton dynamics, ranging from exciton-exciton scattering~\cite{schultheis1986ultrafast, wegener1990line, leo1990effects, lindberg1992theory} to exciton-phonon coupling~\cite{carter2007echo, jakubczyk2016impact}, and it has been applied to three dimensional bulk~\cite{schultheis1986ultrafast, becker1988femtosecond, lohner1993coherent}, two dimensional quantum well~\cite{webb1991observation, langer2014access}, one dimensional nanowire~\cite{wagner2017population}, and zero dimensional quantum dot structures~\cite{langbein2005microscopic, moody2013fifth, delmonte2017coherent, kosarev2020accurate}. However, the technique is not restricted to solid state samples; it has also been applied to liquids~\cite{vohringer1995time, de1998ultrafast, jordanides1999solvation}. In its classical form the photon echo is based on a four-wave mixing (FWM) process and therefore it constitutes a nonlinear process of third order ($\chi^{(3)}$) in the low excitation limit~\cite{cho1992photon}. Especially in single low-dimensional systems like quantum dots it requires large effort to detect such nonlinear optical signals~\cite{cundiff2008coherent, langbein2010coherent}. Interestingly, due to their strong excitonic optical response, monolayers of transition metal dichalcogenides (TMDCs) show a remarkable signal strength in FWM spectroscopy~\cite{moody2015intrinsic, jakubczyk2016radiatively, hao2017neutral, jakubczyk2019coherence}. So far the direct correspondence between inhomogeneous spectral broadening and photon echo duration has been used to map the inhomogeneity of TMDC monolayers~\cite{moody2015intrinsic, jakubczyk2016radiatively, jakubczyk2019coherence, boule2020cohe}.

We here exploit the strong excitonic optical response of TMDC monolayers, and explore six-wave mixing (SWM) signals from a MoSe$_2$ monolayer in the low excitation limit, which here represents the $\chi^{(5)}$-regime~\cite{tominaga1995fifth, bolton2000demonstration, voss2002biexcitonic, zhang2007opening}. We find that the signal dynamics exhibit a characteristic temporary suppression depending on the delay between the pulses. Supported by a theoretical model based on the local field effect describing exciton-exciton interaction, we explain that this suppression can be understood as the formation of a {\it destructive photon echo}.

\section{Results}
\subsection{Sample and four-wave mixing dynamics}\label{sec:2}
\begin{figure}[h]
	\centering
	\includegraphics[width=\columnwidth]{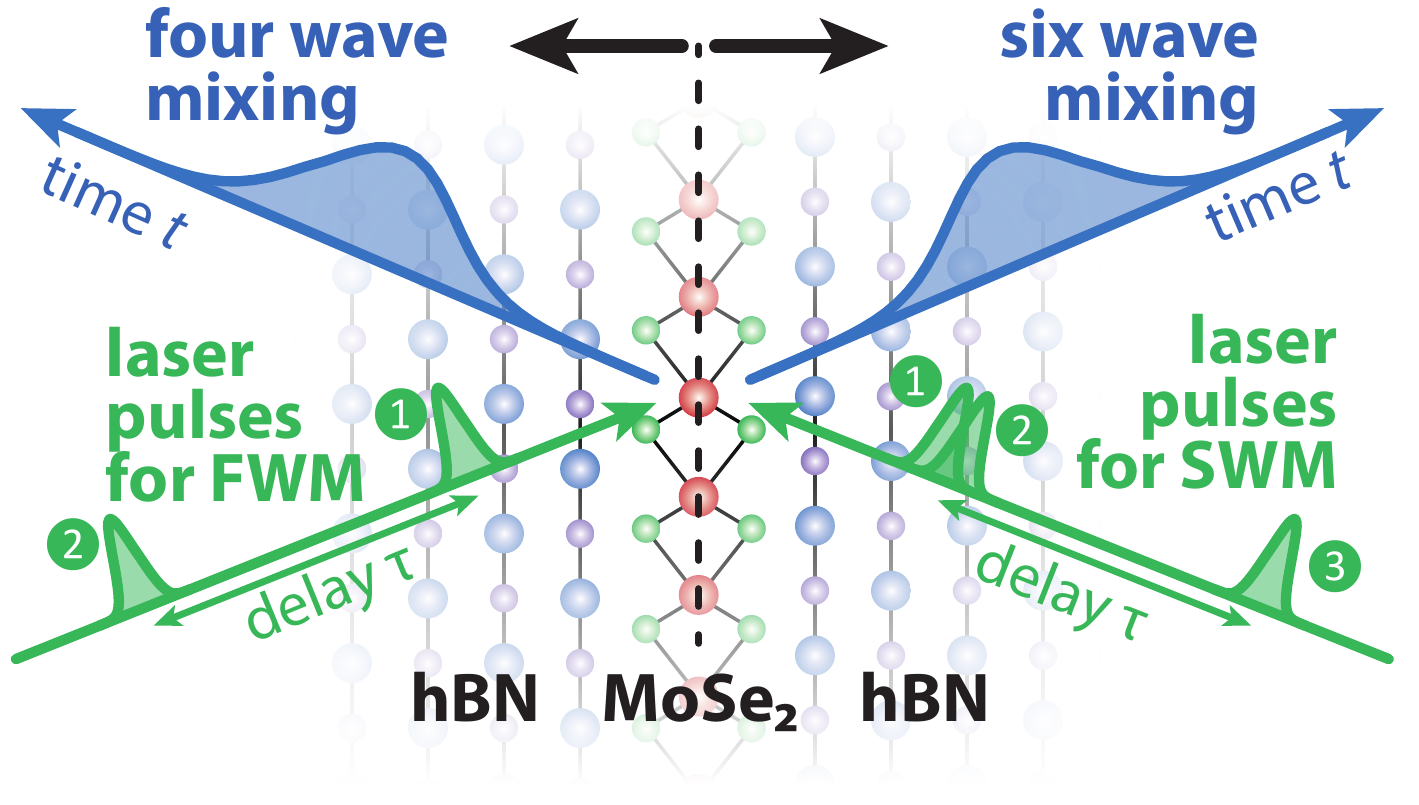}
	\caption{Sample structure consisting of a multilayer hBN, monolayer MoSe$_2$, multilayer hBN stack. (Left) Four-wave mixing (FWM) generated by two laser pulses with tunable delay $\tau$. (Right) Six-wave mixing (SWM) with three excitation pulses having a tunable delay $\tau$ between pulses 2 and 3. The FWM and SWM dynamics are measured in real time $t$.}
	\label{fig:sample}
\end{figure}%

\begin{figure}[t]
	\centering
	\includegraphics[width=\columnwidth]{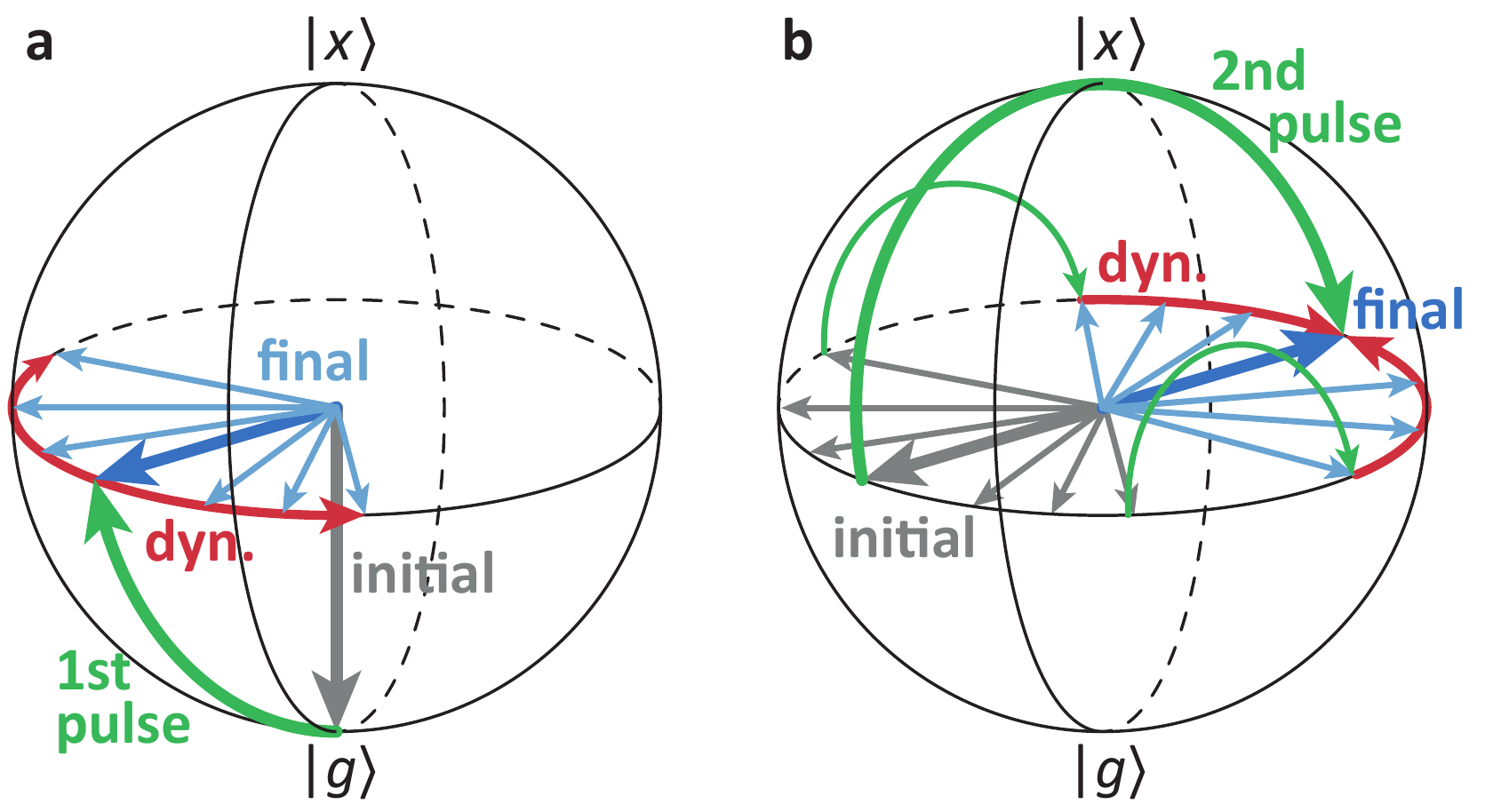}
	\caption{Schematic Bloch vector image of the photon echo process. (a) 1st laser pulse excitation with following dissipation. (b) 2nd excitation inverting all coherences and following rephasing. Laser pulse rotations are depicted in green, initial and final Bloch vectors in grey and blue, respectively, and the inhomogeneity induced coherence dynamics in red.}\label{fig:echo}
\end{figure}

We perform multi-wave mixing experiments on an hBN/MoSe$_2$/hBN heterostructure as schematically depicted in Fig.~\ref{fig:sample}. By radio frequency modulating the incoming beams, the different pulses are labeled by phases $\phi_j$~\cite{langbein2006heterodyne}. After heterodyning the emitted light with a specific $N$-wave mixing (NWM) phase, that is a particular phase combination of the form $\phi_{\rm NWM}=\sum_j a_j\phi_j$ ($a_j\in\mathbb Z$, $\sum_j|a_j|=N-1$, $\sum_ja_j= 1$), we retrieve the corresponding nonlinear signal from the investigated monolayer~\cite{mukamel1999principles}. Details are given in the experimental section. We use the same sample as investigated in Ref.~\cite{boule2020cohe} where the echo formation in two-pulse FWM signals with $\phi_{\rm FWM}=2\phi_2-\phi_1$ was used to study the inhomogeneity of the structure. The well known photon echo appears due to the dephasing impact of the structural inhomogeneity on the exciton's coherence as schematically depicted via Bloch vectors in Fig.~\ref{fig:echo}~\cite{shah2013ultrafast}. For the sake of simplicity in the illustration we show a combination of a $\pi/2$ and a $\pi$ pulse, which results in a photon echo in the full coherence. However, when considering the FWM coherence characterized by the phase $\phi_{\rm FWM}$ the photon echo appears for any combination of pulse areas. The first laser pulse in Fig.~\ref{fig:echo}(a) having a pulse area of $\theta_1=\pi/2$ and arriving at the time $t=-\tau$ generates an exciton coherence. Because of the presence of different transition energies, originating from strain and dielectric variations, the coherences generated by the first laser pulse oscillate with different frequencies resulting in a dephasing of the total coherence. As depicted in Fig.~\ref{fig:echo}(b), after a delay $\tau$, i.e., at the time $t=0$, a second laser pulse with the pulse area $\theta_2=\pi$ inverts all Bloch vectors. This is followed by a rephasing of the different coherences. The rephasing takes the same time as the dephasing and therefore the FWM signal is significantly enhanced due to constructive interferences at the time precisely given by the delay time $t=\tau$ between the two pulses. 
\begin{figure}[h]
	\centering
	\includegraphics[width=\columnwidth]{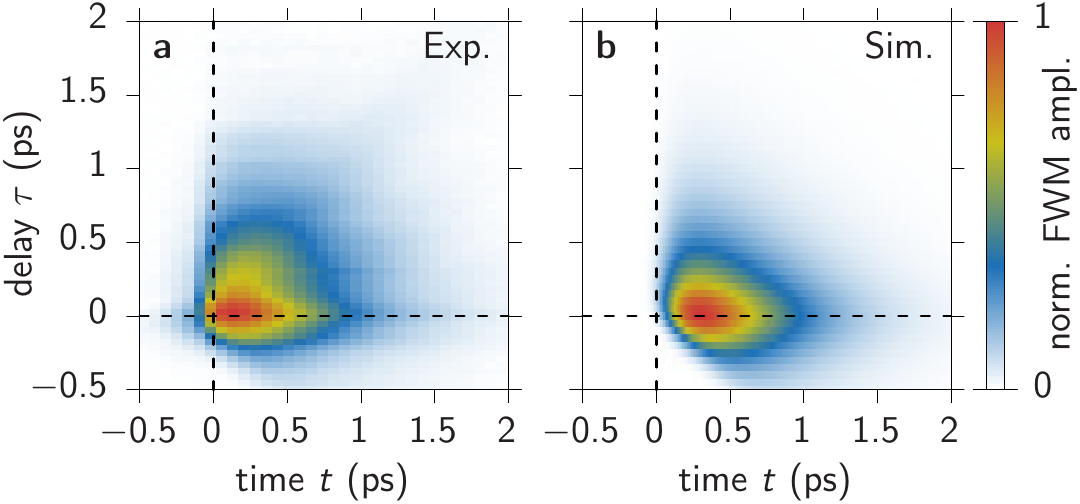}
	\caption{Four-wave mixing dynamics as a function of real time $t$ and delay $\tau$. (a)~Experiment and (b) simulation.}\label{fig:no_echo}
\end{figure}%

In Fig. \ref{fig:no_echo}(a) we show the measured FWM signal from our sample as a function of the real time $t$  after the second pulse and the delay $\tau$. The pulse alignment is such that the pulse with phase $\phi_1$ arrives at $t=-\tau$ and the pulse with $\phi_2$ at $t=0$. We find that the amplitude is not concentrated along the diagonal, which would represent a photon echo. Instead, starting from its maximum around $(\tau,t)=(0,0)$ it basically decays to positive delays and times. This demonstrates that the investigated sample position is virtually homogenous because the photon echo is not present. Figure~\ref{fig:no_echo}(b) shows the corresponding simulated FWM amplitude $\left| S_{\rm FWM}\right|$ within a few-level system where we take the exciton transfer into the optically uncoupled valley as well as the exciton-exciton interaction in terms of the local field model into account. It describes the optically induced dynamics of the exciton's coherence $p$ and its occupations $n$ and $n'$ via~\cite{wegener1990line, hahn2021infl, rodek2021local}

\begin{subequations}\label{eq:3LS}\begin{align}
	\frac{{\rm d} p}{{\rm d} t} &= i(1-2n)[\Omega(t) + V p]  -  ( \beta + i\omega_0 ) p \,, \\
	\frac{{\rm d} n}{{\rm d} t} &= 2{\rm Im}[\Omega^*(t)p]- \Gamma n - \lambda (n - n')\,,\\
	\frac{{\rm d} n'}{{\rm d} t} &= - \Gamma n' - \lambda (n' - n)\,.
\end{align}\end{subequations}
Here, $n$ and $n'$ are the occupations in the optically coupled and uncoupled valleys, respectively, $\Omega(t)$ is the time dependent Rabi frequency describing the optical excitation by co-circlarly polarized pulses, $\beta$ and $\Gamma$ are the dephasing and the decay rate, respectively, $\hbar\omega_0$ is the exciton energy in the absence of the local field coupling, and $V$ is the strength of the local field coupling. Compared to Ref.~\cite{rodek2021local} we do not find a significant impact of excitation induced dephasing as discussed in more detail in the supporting information (SI).  As recently studied in Ref.~\cite{rodek2021local} we additionally take an intervalley scattering contribution with the rate $\lambda$ into account. In the special case $\lambda=0$, the system reduces to a two-level system. We have recently investigated the local field model in the context of nonlinear optical signals focussing on FWM~\cite{hahn2021infl} and pump-probe spectroscopy~\cite{rodek2021local}, showing that the handy description produces resonant optical spectroscopy signals that are consistent with experiment. In a nutshell, the local field effect leads to energy shifts of the exciton depending on the its occupation. In the limit of ultrafast optical pulses Eqs.~\eqref{eq:3LS} can be solved analytically relating the coherence $p^+$ and the occupation $n^+$ after the pulse to the respective values $p^-$ and $n^-$ immediately before the pulse
\begin{subequations}\label{eq:pulse}\begin{align}
	p^+ &= p^- \cos^2\left(\frac{\theta}{2}\right) + \frac{i}{2}\sin(\theta)(1-2n^-) e^{i\phi} \notag\\
		&\qquad + \sin^2\left(\frac{\theta}{2}\right) p^{-*} e^{i2\phi}\notag\\
		&\approx p^- + i \frac{\theta}{2} (1-2n^-) e^{i\phi}\, ,\\
	n^+ &= n^- + \sin^2\left(\frac{\theta}{2}\right) (1-2n^-) +\sin(\theta){\rm Im}\left(p^-e^{-i\phi}\right) \notag\\
		&\approx n^- + \theta\,{\rm Im}\left(p^-e^{-i\phi}\right) \, ,
\end{align}\end{subequations}
where $\theta$ is the pulse area and $\phi$ its phase. The approximations in Eqs.~\eqref{eq:pulse} describe the light-field induced contributions of a single pulse in first order of the pulse area, which is sufficient for reproducing the contributions of the order $V^2$ to the SWM signal, discussed in the main text of this work (other contributions, involving terms up to the second order in the pulse areas, are discussed in the SI.

Starting from the excitonic ground state characterized by $n=n_1^-=0$, $p=p_1^-=0$ the first pulse in the linear order creates
\begin{subequations}\label{eq:1_plus}\begin{align}
	p_1^+ &\approx  i \frac{\theta_1}{2} e^{i\phi_1}\, ,\label{eq:1_plus_p}\\
	n_1^+  &\approx  0 \, ,
\end{align}\end{subequations}
which shows that relevant exciton occupations will only be created by second laser pulse from $p_1^+$. Our goal is to derive the nonlinear optical response in the lowest order of the pulse area $\theta$ because the experiments are performed with low pulse powers. In the case of the SWM signal this is the fifth order $\mathcal O(\theta^5)$ ($\chi^{(5)}$-regime).

Once optical pulses have generated coherences $p_0$ and occupations $n_0$ the corresponding free propagation [$\Omega~=~0$, in Eqs.~\eqref{eq:3LS}] is governed by pure dephasing of $p$, exciton decay and intervalley scattering of $n^{(\prime)}$, and local field coupling between $p$ and $n$. The corresponding dynamics can also be calculated analytically. Focusing on the optically addressed occupation first, its time-dependence reads
\begin{subequations}\begin{align}
	n(t) &= \frac{n_0+n_0'}{2}e^{-\Gamma t} + \frac{n_0-n_0'}{2}e^{-(\Gamma +2\lambda)t} \,,
\end{align}
which leads to a balanced occupation between the two valleys $n$ and $n'$ on the timescale $1/(2\lambda)$ and a decay of both occupations with the rate $\Gamma$. As known from literature \cite{schmidt2016ultrafast, wang2018colloquium, rodek2021local} and as considered in this work the intervalley scattering is typically significantly faster than the decay, i.e., $\Gamma \ll 2\lambda$. As a simplifying approximation for the sake of interpreting the results we can therefore assume that the occupations are balanced rapidly after an optical pulse, i.e., 
\begin{align}\label{eq:scatter_approx}
	n(t) &\approx \frac{n_0+n_0'}{2}e^{-\Gamma t} \,.
\end{align}
This step might lead to slight deviations for short delays in the range $\tau\approx1/(2\lambda)$ which are however hardly visible for the parameters chosen here as shown in the SI. Further we can only optically manipulate the occupation $n$, while $n'$ remains unchanged by the applied laser pulses. According to Eq.~\eqref{eq:1_plus_p} these pulses moreover add phase labels $\phi$. As explained at the beginning of this section and as practically applied below, we are only interested in specific phase combinations that describe the considered wave-mixing signal. Consequently, any change of the occupation $n$ is labeled by phase factors which do not apply to the other valley $n'$. Therefore, the latter is irrelevant for the final optical signal and we can consider $n_0'=0$ in Eq.~\eqref{eq:scatter_approx} resulting in
\begin{align}\label{eq:scatter_approx_2}
	n(t) &\approx \frac{n_0}{2}e^{-\Gamma t} \,.
\end{align}\end{subequations}
We want to remark that in the opposite limit of a very slow scattering rate $2\lambda\ll \Gamma$ the occupation dynamics would directly be given by $n(t)=n_0e^{-\Gamma t}$. In this case all later discussions would work in the same way. By replacing $V\to 2V$ in all following derivations one can even directly retrieve the corresponding equations.

Based on this approximation for the occupation dynamics we can calculate the coherence dynamics
\begin{align}
	p(t) &= p_0 \exp\left[ -i\frac{2V}{\Gamma}\int\limits_0^t {\rm d}t' \,n(t')\right] e^{(iV-\beta )t} \notag\\
		&\approx p_0 \exp\left[ -i\frac{2V}{\Gamma}\frac{n_0}{2}\left(1-e^{-\Gamma t}\right)\right] e^{(iV-\beta )t} \notag\\
		&\approx p_0 e^{ iVn_0t } e^{(iV-\beta )t}  \notag\\
		&\approx p_0 \left[ 1 - iVn_0 t  - \frac12(Vn_0t)^2\right] e^{(iV-\beta) t}\,. \label{eq:mixing}
\end{align}
In the first approximation step we have used the approximated occuption from Eq.~\eqref{eq:scatter_approx_2}. As the exciton decay is much slower than the dephasing $\Gamma \ll \beta$, in the second step we take $\Gamma\to 0$. Finally, we keep the local field-induced contributions up to the second order in the exciton occupation, i.e., considering $Vn_0t\ll 1$. Below we will see that these are the terms which contribute to the SWM signal in the $\chi^{(5)}$-regime. In the SI we show that only terms up to $\mathcal O(V^2)$ contribute to the $\chi^{(5)}$-regime of the SWM signal, while contributions with higher powers in $V$ appear in higher orders of the optical field. The last equation tells us that the local field induced mixing of the occupation $n_0$ with the coherence $p_0$ into the coherence $p(t)$ in the lowest order grows linearly in time with a rate given by $n_0$ and the local field strength $V$. We will use this argument later on.

We simulate the FWM signal with the phase combination $2\phi_2-\phi_1$ by calculating the coherence dynamics $p(t)$ following a two-pulse sequence and filter this quantity with respect to the required phase combination as described in \linebreak Refs.~\cite{hahn2021infl, rodek2021local}. With this we find the signal dynamics $|S_{\rm FWM}|$\linebreak$\sim \left|^{2\phi_2-\phi_1}p_2(t,\tau)\right|$ depicted in Fig.~\ref{fig:no_echo}(b). To achieve the excellent agreement with the experiment in Fig.~\ref{fig:no_echo}(a) we used a local field strength of $V=100$~ps$^{-1}$ and pulse areas of $\theta_1=\theta_2/2=\theta=0.02\pi$, a Gaussian pulse duration of $\Delta t=70$~fs (standard deviation), a dephasing rate of $\beta=3$~ps$^{-1}$, an intervalley scattering rate $\lambda=4$~ps$^{-1}$, and a decay rate of $\Gamma=0.6$~ps$^{-1}$. Note, that the depicted signal was calculated numerically because we considered a non-vanishing pulse duration and we did not employ the approximations mentioned in Eqs.~\eqref{eq:pulse} -- \eqref{eq:mixing}. To set the strength of the local field coupling into context, typical values for GaAs quantum wells were reported in the range of a few meV~\cite{kim1992unusually, mayer1994evidence}, which is at least one order of magnitude smaller than considered here.

\subsection{Six-wave mixing dynamics}\label{sec:SWM}
\begin{figure}[t]
	\centering
	\includegraphics[width=\columnwidth]{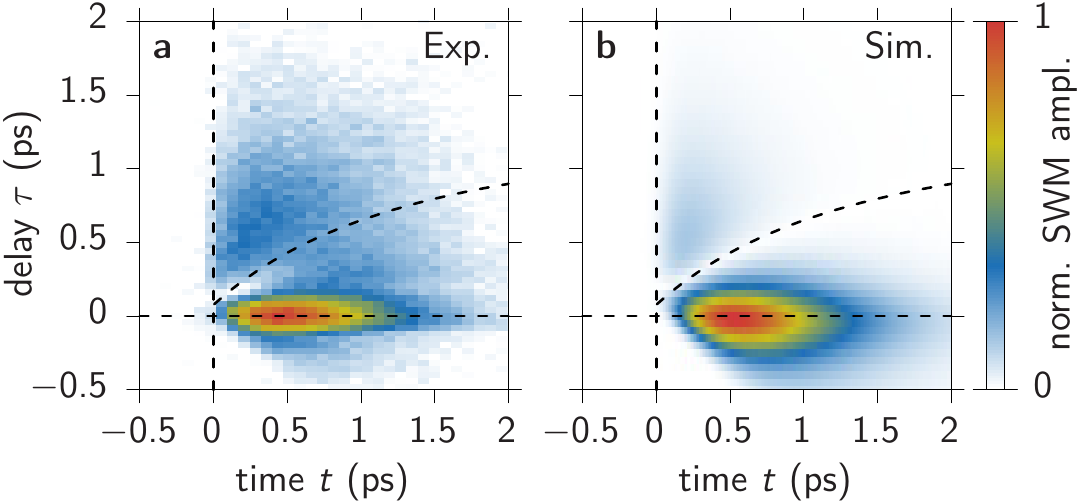}
	\caption{Six-wave mixing dynamics as a function of real time $t$ and delay $\tau$. (a) Experiment, (b) simulation, with the curved dashed line depicting Eq.~\eqref{eq:minimum}.}\label{fig:antiecho}
\end{figure}

In the present study we go one step further in multi-wave mixing and consider one of the possible SWM signals generated by three laser pulses, namely \linebreak $\phi_{\rm SWM} = 2\phi_3-2\phi_2+\phi_1$. In principle there are two delays in this pulse sequence that could be varied but, as schematically shown in Fig.~\ref{fig:sample} (right), we set the delay between the first two pulses to $\tau_{12}=0$ and only vary the second one $\tau_{23}=\tau$. The impact of a non-vanishing $\tau_{12}$ is discussed in the SI. In the pure two-level system this configuration probes the polarization, i.e., it contains the same information as the FWM signal discussed before as shown in the SI. The measured SWM dynamics as a function of the real time $t$ after the third pulse and the delay $\tau$ are shown in Fig.~\ref{fig:antiecho}(a). Here, the two pulses with $\phi_1$ and $\phi_2$ arrive at $t=-\tau$ and the pulse with $\phi_3$ at $t=0$. The signal consists of a strong maximum at small $t\approx0.5\,$ps and $\tau\approx 0$. Moving to negative delays $\tau<0$ (pulse 3 is arriving before 1 and 2), the signal is strongly damped. For positive delays $\tau>0$ it decays much slower on the same timescale as the FWM signal in Fig.~\ref{fig:no_echo}. We find a remarkable depression of the signal that stretches along a curved diagonal given by Eq.~\eqref{eq:minimum} (dashed black line), as will be derived below. In correspondence with the previously described constructive signal enhancement in the photon echo, we call this pronounced signal reduction a \textit{destructive photon echo}. Later, we will discuss criteria justifying the labeling of this feature as an echo effect.

To identify the origin of this peculiar dynamical feature we model the SWM signal within the local field model described above. We consider the same system parameters as for the FWM signal but choose equal pulse area for all three pulses $\theta_1=\theta_2=\theta_3=\theta=0.02\pi$, in agreement with the experiment. In the SI it is shown that the exact choice of $\theta$ in the low excitation regime does not change the SWM signal dynamics. The SWM signal is extracted via the phase combination $2\phi_3-2\phi_2+\phi_1$ and the signal dynamics $\left| S_{\rm SWM}\right| \sim \left|^{2\phi_3-2\phi_2+\phi_1}p_3(t,\tau)\right|$ are depicted in Fig.~\ref{fig:antiecho}(b). The retrieved signal agrees very well with the respective experiment in Fig.~\ref{fig:antiecho}(a) showing the same characteristic suppression of the signal, i.e., the destructive photon echo. We give an overview regarding the impact of the different parameters on the SWM signal in the SI. Most importantly, it is shown in the SI, that the SWM dynamics do not exhibit the suppression for small local field strengths~$V$ demonstrating that this feature is a result of exciton-exciton interaction. Other specific features in the signal's dynamics like the asymmetric decay between positive and negative delays or between $\tau$ and $t$ were studied in detail in Ref.~\cite{hahn2021infl} and behave similarly in SWM. The destructive photon echo effect should also be present in an inhomogeneously broadened system, where also a traditional constructive photon echo develops. As the constructive echo selects only a specific time interval of the emitted signal around $t=\tau$, it leads to a suppression of the entire SWM signal for all other times. Consequently, the SWM signal and therefore the destructive photon echo, which bends away from the diagonal (discussed later), is only visible in the vicinity of $t=\tau$. This aspect is discussed in more detail in the SI.

\begin{figure}[t]
	\centering
	\includegraphics[width=\columnwidth]{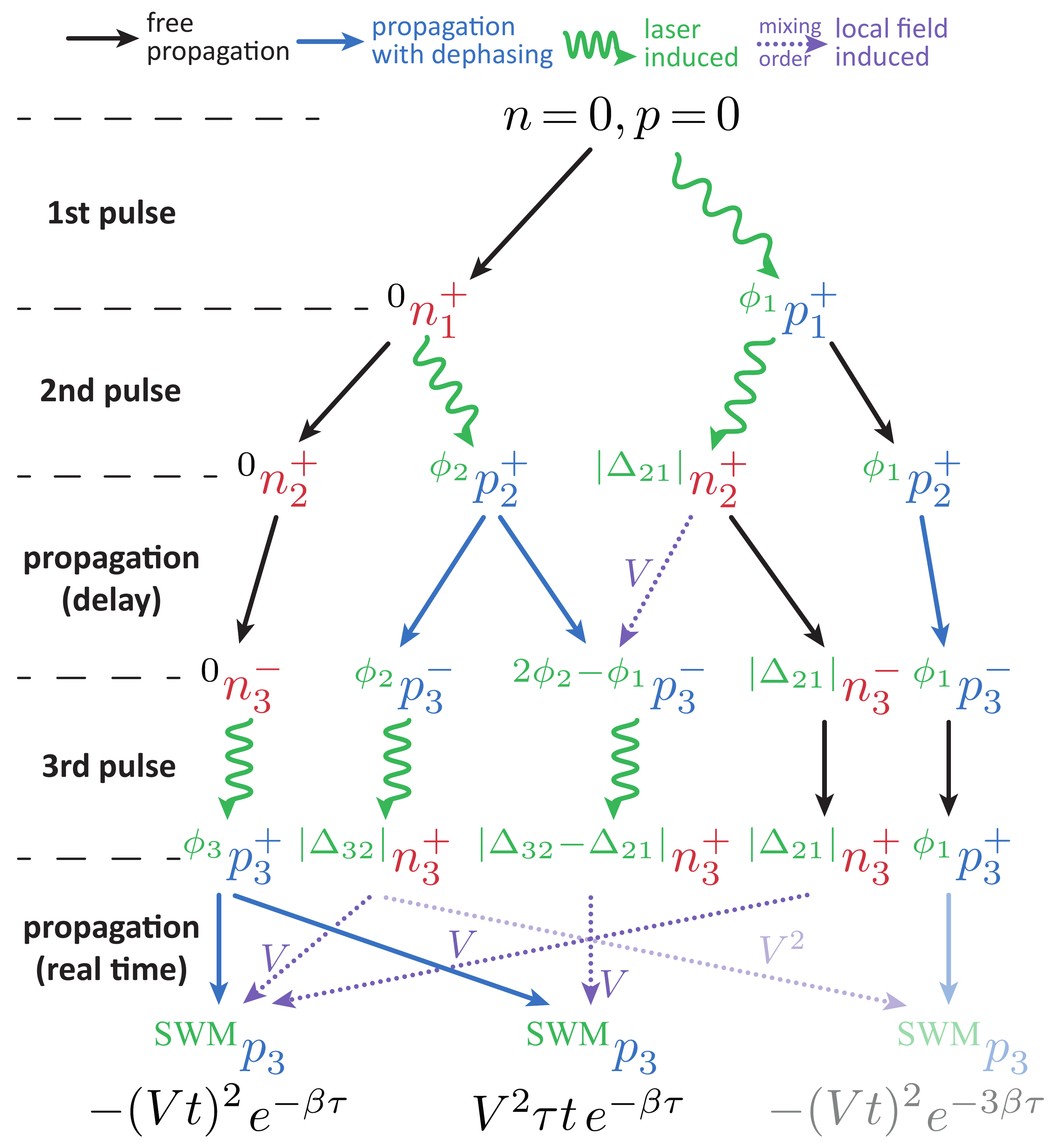}
	\caption{Flow chart for the three main contributions to the SWM signal listing intermediate phase-filtered coherences ($p$, blue) and occupations ($n$, red). Green waved arrows show pulse induced, violet dotted ones local field induced, black ones free dynamics without, and blue ones with dephasing. The phase differences are defined as $\Delta_{nm}=\phi_n-\phi_m$. The flow chart holds for any order of the optical field. However, in the linear response regime considered here it is $^0n_1^+={}^0n_2^+={}^0n_3^-=0$. }\label{fig:flow}
\end{figure}%

Note, that the simulation shown in Fig.~\ref{fig:antiecho}(b) takes the non-vanishing pulse duration into account and is therefore performed numerically. In the limit of ultrafast laser pulses we can find analytical expressions for the SWM signal. Given that the experiment is carried out with weak pulse powers, we restrict the following studies on the lowest order in the optical field which is the $\chi^{(5)}$-regime. In this order we have already eight different contributions as derived in the SI. From these we will focus on the ones with the strongest local field contribution which is $\mathcal O(V^2)$, i.e., we omit all terms $\mathcal O(V^1)$ and $\mathcal O(V^0)$. The reason for this is the absence of the destructive echo for small $V$. In Ref.~\cite{hahn2021infl} we have derived a flow chart representation for the construction of nonlinear wave mixing signals. In Fig.~\ref{fig:flow} we employ this procedure to disentangle the origin of the different contributions to the SWM signal. The flow chart only shows coherences $p$ (blue) and occupations $n$ (red) with phase combinations (green, given as left indices) relevant for the final signal; corresponding flow charts for the contributions with $\mathcal O(V^1)$ and $\mathcal O(V^0)$ are provided in the SI. Conveniently, the $\mathcal O(V^2)$-contributions can be derived with the appoximations given in Eq. \eqref{eq:pulse}. Note, that we introduce phase differences $\Delta_{nm}=\phi_n-\phi_m$ here. The right lower index refers to the pulse number, while the upper $-$ ($+$) indicates times immediately before (after) that pulse. Starting from the excitonic ground state with $n=0$, $p=0$ the first pulse generates the occupation $^0n_1^+$ and the coherence $^{\phi_1}p_1^+$. Note that in the scheme we do not restrict ourselves to the lowest order contributions in the light field and thus include also the occupation which is of second order in the pulse amplitude. The second pulse arrives at the same time ($\tau_{12}=0$) and creates two occupations $^0n_2^+$ and $^{|\Delta_{21}|}n_2^+$ and the coherence $^{\phi_2}p_2^+$. During the following propagation for the time of the delay $\tau$ two relevant things happen: On the one hand all coherences experience dephasing (blue arrows). On the other hand $^{\phi_2}p_2^+$ is mixed with $^{|\Delta_{21}|}n_2^+$ via the local field coupling resulting in the coherence $^{2\phi_2-\phi_1}p_3^-$ before the third pulse. Note, that this contribution carries the FWM phase $\phi_\text{FWM} = 2\phi_2-\phi_1$ which we will come back to below. After the final third pulse we have five relevant terms~\footnote{Note, that the contribution to ${}^{2\phi_2-\phi_1}p_3^-$ generated directly by the first and second pulse is of lower order in $V$ and therefore neglected here but considered in the SI. However, it is fully included in the numerical solution.}: The coherence $^{\phi_3}p_3^+$ and the three occupations $^{|\Delta_{32}|}n_3^+$, $^{|\Delta_{32}-\Delta_{21}|}n_3^+$, and $^{|\Delta_{21}|}n_3^+$. The polarization ${}^{\phi_1}p_1^+$ is not affected by the second and third pulse and just evolves into ${}^{\phi_1}p_3^+$. During the remaining propagation step in real time $t$ the three relevant SWM contributions are generated by local field mixing processes according to Eq.~\eqref{eq:mixing}.

Considering the contribution on the right first, we have to mix $^{\phi_1}p_3^+$ with $^{|\Delta_{32}|}n_3^+$ twice. This results in the phase combination $\phi_1+2(\phi_3-\phi_2)$ which is the SWM phase combination. According to Eq.~\eqref{eq:mixing} each of these local field mixing processes contributes with an amplitude of $Vt$ resulting in the amplitude $(Vt)^2$. In addition the amplitude is damped due to the dephasing happening during the delay. Following the two paths in the diagram back to this propagation step we find that $^{\phi_1}p_2^+ \,\to\, ^{\phi_1}p_3^-$ and $^{\phi_2}p_2^+ \,\to\, ^{\phi_2}p_3^-$ contribute with a dephasing term $\sim~\!\!e^{-\beta \tau}$. The latter one is used twice due to the double local field mixing resulting in the total damping rate of $e^{-3\beta \tau}$.

Moving on to the left contribution in Fig.~\ref{fig:flow} we find that in the last propagation the coherence $^{\phi_3}p_3^+$ is local-field mixed with $^{|\Delta_{32}|}n_3 ^+$ and $^{|\Delta_{21}|}n_3 ^+$ once, resulting in the SWM phase combination $\phi_3+(\phi_3-\phi_2)-(\phi_2-\phi_1)$. Each of these processes contributes an amplitude of $Vt$ again resulting in $(Vt)^2$. The difference to the first term is that here only  $^{|\Delta_{32}|}n_3 ^+$ stems from a coherence, namely $^{\phi_2}p_2^+$ which experiences a dephasing during the delay. Consequently, the entire contribution is damped by $e^{-\beta \tau}$. This already shows that this contribution is more important for the final SWM signal than the right term discussed first.

The final main contribution is the middle one in Fig.~\ref{fig:flow}. Here, in the last real time propagation the local field mixing happens between $^{\phi_3}p_3^+$ and $^{|\Delta_{32}-\Delta_{21}|}n_3^+$ which contributes with an amplitude of $Vt$. Following the path back, we find that the occupation is created from the polarization $^{2\phi_2-\phi_1}p_3^-$ which itself is produced by a local field mixing step between $^{\phi_2}p_2^+$ and $^{|\Delta_{21}|}n_2^+$. The mixing process lasts for the delay time and therefore contributes with a factor $V\tau$ to the final SWM amplitude. This is also the step where the only dephasing happens resulting in a final amplitude of $V^2\tau t e^{-\beta \tau}$.

\begin{figure}[t]
	\centering
	\includegraphics[width=\columnwidth]{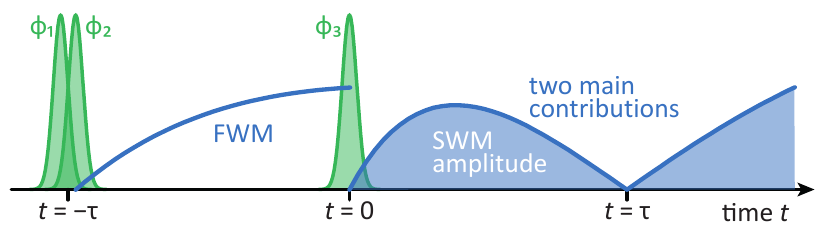}
	\caption{Schematic picture of the pulse sequence and signal development from the two main contributions.}\label{fig:easy_signal}
\end{figure}

Directly comparing the three contributions in Fig.~\ref{fig:flow} we find that the right one is smaller than the other two due to the stronger dephasing during the delay. We will therefore disregard this term from now on. The other two terms are of the order $V^2$ and exhibit the same dephasing with $e^{-\beta \tau}$. From the full derivation given in the SI we find that these two terms in Fig.~\ref{fig:flow} carry opposite signs. We will explain these signs later when introducing an effective Bloch vector picture to illustrate the destructive echo formation. Finally, adding the two terms we get
\begin{align}\label{eq:swm}
	^{\rm SWM}p_3(t,\tau) \approx V^2(\tau t-t^2)e^{-\beta(\tau+t)}\,,
\end{align}
where we also added the dephasing rate for the propagation in $t$ after the last pulse. We find that the two terms exactly compensate each other for $t=\tau$. This renders our first step towards the understanding of the destructive echo. We identified two paths (signal contributions) with the same damping but with different magnitudes depending on the delay $\tau$ that act destructively in the total SWM signal. To illustrate the interplay between these two terms in Fig.~\ref{fig:easy_signal} we schematically plot the applied laser pulses and the absolute value of the final SWM signal retrieved from the previous derivations. We also include the intermediate growth of the FWM coherence lasting for the delay $\tau$.

The full SWM signal in the $\chi^{(5)}$-regime is derived in the SI and reads
\begin{align}\label{eq:swm_analyt}
	 ^{\rm SWM}p_3(t,\tau) &= \left(\frac{\theta}{2}\right)^5 \bigg[ i + V( \tau - 4t ) - \frac i2(Vt)^2 e^{-2\beta\tau} \notag\\
	 &\qquad+ i V^2t(\tau-t)  \bigg] e^{-\beta(t+\tau)}\,.
\end{align}
Interestingly, we also find a suppression of the signal when only considering the terms $\mathcal O(V^1)$, namely for $t=\tau/4$. We obviously do not find such a feature in our measurement, which shows that the linear oder in $V$ does not have a significant contribution.

\begin{figure}[h]
	\centering
	\includegraphics[width=\columnwidth]{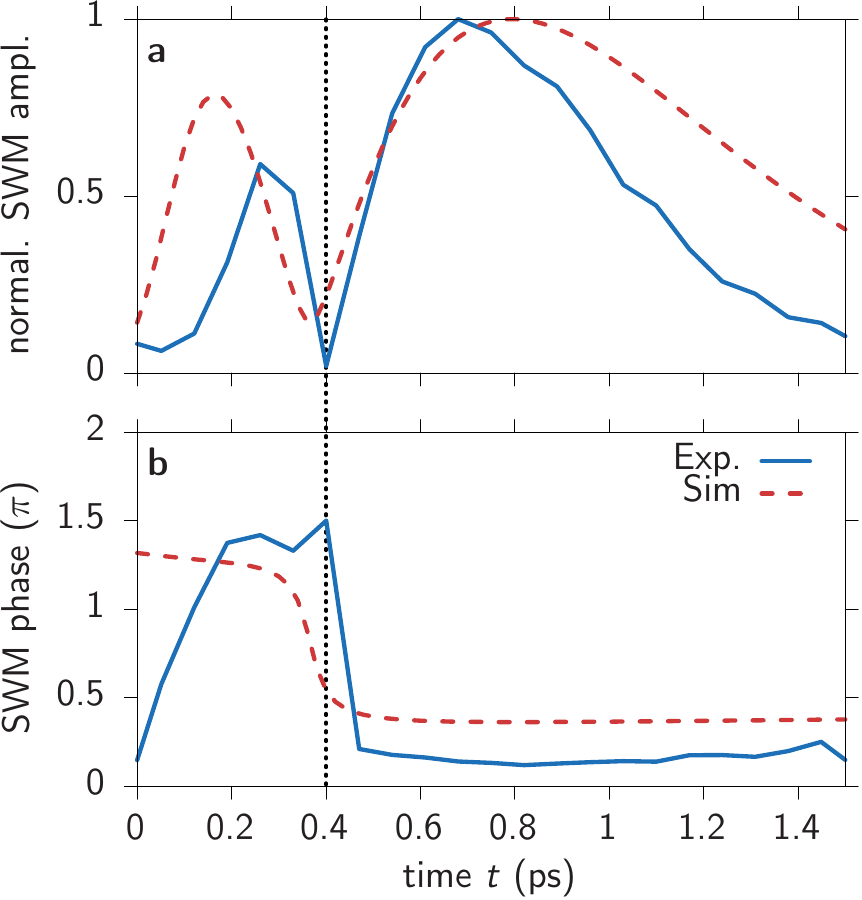}
	\caption{Amplitude and phase dynamics of the destructive echo. (a) SWM amplitude dynamics for a delay of $\tau=0.35$~ps. Experiment as solid blue and simulation as dashed red line. (b)~Respective dynamics of the phase of the SWM signal.}\label{fig:phase}
\end{figure}
One advantage of the spectral interferometry in the applied approach is the possibility to detect -- besides the amplitude -- also the phase of the SWM signal~\cite{kasprzak2011coherent}. We see that for $t<\tau$ the positive contribution $ \sim \tau t$ dominates while for $t>\tau$ the negative one $\sim - t^2$ is larger. Therefore we expect a phase jump when crossing the destructive echo in time $t<\tau \to t>\tau$. To confirm this in Fig.~\ref{fig:phase}(a) we plot the measured (solid blue) and calculated (dashed red) SWM signal amplitude as a function of time~$t$ for $\tau=0.35$~ps. We slightly adjusted the pulse duration to $\Delta t=80$~fs to achieve the good agreement with this experiment. The finding that the simulated signal does not drop to zero shows the impact of the non-compensating contributions $\sim V^2$, like the ordinary SWM signal in Eq.~\eqref{eq:swm_analyt}, and the influence of the considered non-vanishing pulse duration. In Fig.~\ref{fig:phase}(b) we show the corresponding SWM phase from $S_{\rm SWM} = \left|S_{\rm SWM}\right|e^{i\varphi}$. We indeed find a jump of approximately $\pi$ at the destructive echo as would be expected for the two dominant terms. However, all other contributions, which are naturally present in the numerical simulation, lead to a reduction of the phase jump and a further distortion of the destructive echo dynamics. We again find that the full depression of the signal happens at $t\approx 0.4$~ps which is slightly later than $t=\tau=0.35$~ps. This slight delay of the effect mainly stems from the non-vanishing decay rate of the exciton $\Gamma\neq 0$.

To include the exciton decay we have to go back to the full dynamics in Eq.~\eqref{eq:mixing} and directly consider\linebreak $V n_0 \left(1-e^{-\Gamma t}\right)/\Gamma\ll 1$ without taking the limit $\Gamma \to 0$ leading to
\begin{subequations}\begin{align}
	n(t) &= n_0e^{-\Gamma t}\,,\\
	p(t) &\approx p_0\left[1- i\frac{V}\Gamma n_0 \left(1-e^{-\Gamma t}\right) - \frac{V^2}{2\Gamma^2}n_0^2 \left(1-e^{-\Gamma t}\right)^2\right].
\end{align}\end{subequations}
Now we consider the different paths in Fig.~\ref{fig:flow} to determine the decay's impact on the different SWM contributions. The term $\sim t^2$ stems from the delay step $^{|\Delta_{21}|}n_2^+ \,\to\, ^{|\Delta_{21}|}n_3^-$, contributing a factor $e^{-\Gamma\tau}$, and the final double wave mixing step gives $\left(1-e^{-\Gamma t}\right)^2$. The $\sim \tau t$ contribution has a local-field mixing step during the delay leading to $\left(1-e^{-\Gamma \tau}\right)$ and only a single local field mixing after the last pulse, i.e., $\left(1-e^{-\Gamma t}\right)$. Finally, the two contributions compensate each other for
\begin{gather}
	\left(1-e^{-\Gamma t}\right)^2e^{-\Gamma \tau} = \left(1-e^{-\Gamma t}\right)\left(1-e^{-\Gamma\tau}\right)\,, \notag\\
	\Rightarrow\tau = \frac 1\Gamma \ln\left(2-e^{-\Gamma t}\right)\, . \label{eq:minimum}
\end{gather}
We find that the decay additionally acts on the local field mixing processes because it dynamically reduces the occupation during the mixing. This affects the $t^2$-contribution in a different way than the part $\sim \tau t$ and the destructive echo gets delayed with respect to $t = \tau$. Equation~\eqref{eq:minimum} does not depend on $V$ as confirmed in the SI. Consequently, it also holds in both limiting cases of small and large intervalley scattering as discussed below Eq.~\eqref{eq:scatter_approx_2} and can be expected to be correct also for intermediate $\lambda$-values. Note, that the curve describing the dynamics of the SWM signal depression in Fig.~\ref{fig:antiecho} follows Eq.~\eqref{eq:minimum} and is additionally shifted by the pulse width $\Delta t$ to shorter times to compensate the non-vanishing pulse duration. The curve almost perfectly follows the distribution of the destructive echo.

\subsection{Quasi-Bloch vector and phase inhomogeneity}\label{sec:Bloch}
As discussed in the previous section, the FWM-polarization constitutes the basis for the SWM signal. Therefore, it is instructive to introduce the concept of the quasi-Bloch vector to explain the wave-mixing origin for the FWM signal in the next section. Building on this picture we will be able to explain the destructive photon echo formation in the SWM signal in the following section.
\subsubsection{Four-wave mixing quasi-Bloch vector}\label{sec:QBV_FWM}
\begin{figure}[h]
	\centering
	\includegraphics[width=\columnwidth]{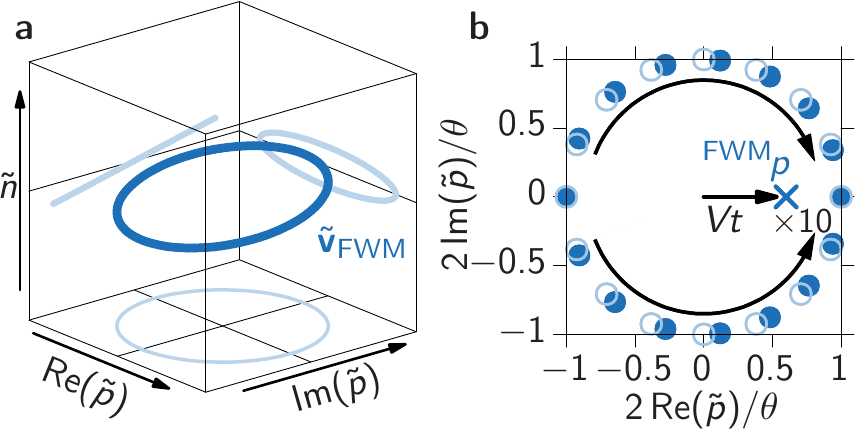}
	\caption{Quasi-Bloch vector of the FWM contribution. (a)~Lissajous curve of the initial distribution of quasi-Bloch vectors. (b) Dynamics of the quasi-Bloch vectors. Initial distribution as bright open circles and for a time $t$~$>$~$0$ after the second pulse as filled dark circles. The final integrated FWM coherence $^{\rm FWM}p$ is marked as blue cross and its propagation is given by $Vt$.}\label{fig:Bloch_FWM}
\end{figure}

The ordinary echo in FWM is most instructively visualized within the Bloch vector picture in Fig.~\ref{fig:echo} where the inhomogeneity leads to different rotation speeds of the various Bloch vector realizations. In that case each exciton energy is represented by a single Bloch vector. Here, we are dealing with a slightly different situation. The wave mixing experiment is repeated numerous times with successively different phase combinations $(\phi_1,\phi_2, \phi_3)$ for each repetition of the measurement. Therefore we have to represent each run of the experiment, i.e., each phase combination, by a single Bloch vector. To find the realized Bloch vectors we have a look at the coherence and occupation immediately after the second pulse~\cite{hahn2021infl}
\begin{subequations}\begin{align}
	p_2^+ &= i \frac \theta2 \left[  e^{i\phi_1} +  e^{i\phi_2} - \theta^2 e^{i(2\phi_2-\phi_1)}\right] + \mathcal{O}(\theta^5)\,,\\
	n_2^+ &= \frac {\theta^2}2 [1+\cos(\phi_2-\phi_1)] + \mathcal{O}(\theta^4)\,,
\end{align}\end{subequations}
where we consider all terms up to the third order in the optical field ($\chi^{(3)}$-regime) because our measurements are performed with low excitation powers. Following the flow chart in Fig.~\ref{fig:flow} we find that the two terms $^{\phi_2}p_2^+$ and $^{|\Delta_{21}|}n_2^+$ perform local field induced mixing during the following propagation before the third pulse (delay step). Exactly this term was identified as a contribution of the FWM signal in Ref.~\cite{hahn2021infl}. Therefore, we will study this process in more detail. Our goal is to isolate a set of Bloch vectors that allows us to directly extract the final wave mixing signal by integrating over the entire set. To achieve this we have to already filter the coherence with the respective wave mixing phase factor. In this case we are interested in the FWM contribution, which means that we have to consider the phase factor $e^{i\phi_{\rm FWM}}=e^{i(2\phi_2-\phi_1)}$ and consequently the filtered polarization $^{\phi_2}p_2^+\,e^{-i\phi_{\rm FWM}}$. This brings us to the two relevant terms
\begin{subequations}\begin{align}
	&\tilde p = {}^{\phi_2}p_2^+\,e^{-i\phi_{\rm FWM}} = i \frac \theta2 e^{-i(\phi_2-\phi_1)} \,, \\
	&\tilde n = {}^{|\Delta_{21}|}n_2^+ = \frac {\theta^2}2 \cos(\phi_2-\phi_1)\,.
\end{align}\end{subequations}
We use these expressions to define a {\it quasi-Bloch vector} for this FWM contribution via
\begin{equation}\label{eq:QB_FWM}
	\tilde{{\bm v}}_{\rm FWM}(t=0) 
	= \begin{pmatrix}
		{\rm Re}(\tilde{p})\\
		{\rm Im}(\tilde{p})\\
		\tilde{n}
	\end{pmatrix}
	= \frac{\theta}{2} \begin{pmatrix}
		\sin(\Delta_{21})\\
		\sin(\Delta_{21}+\pi/2)\\
		\theta \sin(\Delta_{21}+\pi/2)
	\end{pmatrix}\,,
\end{equation}
with the previously introduced phase difference $\Delta_{21}=\phi_2-\phi_1$. We call this quantity {\it quasi-Bloch vector} because it does not represent the entire density matrix, as the full Bloch vector does. When varying $\Delta_{21}$ as it is done in experiment and simulation, we already see that this set of quasi-Bloch vectors follows a three dimensional Lissajous curve with a frequency relationship of 1:1:1 as depicted in Fig.~\ref{fig:Bloch_FWM}(a). It appears as a circle tilted around the Re($\tilde{p}$)-axis and its projections on the different planes of the coordinate system form Lissajous curves with 1:1 frequency relations. This results in a distribution of occupations $\tilde{n}$, whose spread, represented by the tilt angle, is given by the considered pulse areas.

Starting from this distribution of quasi-Bloch vectors directly after the second laser pulse their dynamics is governed by the local field mixing between $\tilde{p}$ and $\tilde{n}$ described by [see Eq.~\eqref{eq:mixing}]
\begin{equation}\label{eq:LF}
	\frac{{\rm d}\tilde{{\bm v}}}{{\rm d}t} = 
	 \tilde{{\bm v}}\times
	\begin{pmatrix}
		0 \\
		0 \\
		V\tilde{n}
	\end{pmatrix}\,.
\end{equation}
These dynamics are a rotation of the coherence $\tilde{p}$ depending on the occupation $\tilde{n}$. Particularly important is that the rotation flips its direction for opposite signs of $\tilde{n}$. At the same time $\tilde{n}$ remains unaffected. Finally, from the distribution of quasi-Bloch vectors at a given time $\tilde{\bm v}(t)$ we can directly retrieve the final wave mixing contribution by integrating over all possible phase combinations, i.e., over $\Delta_{21}$. We directly see that the integral over $\tilde{n}$ always vanishes and we are left with the final FWM coherence
\begin{equation}\label{eq:Quasi_Bloch}
	^{\rm FWM}p(t) = \int\limits_0^{2\pi}{\rm d}\Delta_{21}\ \tilde{p}(t;\Delta_{21})\,.
\end{equation}

For the initial distribution of quasi-Bloch vectors in\linebreak Eq.~\eqref{eq:QB_FWM} we find that the final FWM coherence vanishes because all vectors are equally distributed on the tilted circle in Fig.~\ref{fig:Bloch_FWM}(a). However, the following dynamics lead to a non-vanishing FWM contribution. The rotation of $\tilde{p}$ with frequencies proportional to $\tilde{n}$ has two crucial consequences: (i) The presence of a distribution of $\tilde{n}$ results in different frequencies. As already discussed in Ref.~\cite{hahn2021infl}  this means that each run of the experiment results in a slightly different emission energy of the exciton and consequently leads to a broadened FWM spectrum. Because the distribution of $\tilde{n}$ stems from the variation of the applied laser pulse phase combinations we call this spectral broadening {\it local field induced phase inhomogeneity}. (ii) According to Eq.~\eqref{eq:LF} the sign of $\tilde{n}$ determines the rotation direction of the quasi-Bloch vectors. The initial distribution of quasi-Bloch vectors in Fig.~\ref{fig:Bloch_FWM}(a) is a circle that is tilted around the Re$(\tilde{p})$-axis and all points on this axis are not affected by the local field induced rotation ($\tilde{n}=0$). This axis remains a symmetry axis also for the dynamics of the quasi-Bloch vectors: one half of the circle rotates clockwise, the other counter-clockwise in exactly the same way. Consequently, when integrating over the relative phase $\Delta_{21}$ the opposing Im$(\tilde{p})$ contributions compensate each other and the final FWM coherence is real, i.e., ${\rm Re}(\tilde{p})\neq 0$. Because only Re($\tilde{p}$) will later contribute to the FWM signal,  the crucial quantity determining the propagation direction of Re($\tilde{p}$) is $\tilde{n}\,{\rm Im}(\tilde{p})$ as can be seen in  Eq.~\eqref{eq:LF}. In Fig.~\ref{fig:Bloch_FWM}(b) we show the quasi-Bloch vector dynamics in the Re($\tilde{p}$),Im($\tilde{p}$)-plane after the second pulse. The bright open circles represent the initial homogeneous distribution and the filled dark circles the distribution for a time $t>0$. Looking at the initial distribution in Fig.~\ref{fig:Bloch_FWM}(a) we find that the semicircle with Im($\tilde{p})<0$ also has $\tilde{n}<0$ and the other semicircle with Im($\tilde{p})>0$ has $\tilde{n}>0$. This means that the former rotate clockwise, while the latter rotate counter-clockwise as indicated by the curved black arrows in Fig.~\ref{fig:Bloch_FWM}(b). Consequently, the weight of all quasi-Bloch vectors, which is depicted by the blue cross, moves into the positive Re($\tilde{p}$)-direction. The speed of this movement is given by $V$ and the tilt angle of the initial quasi-Bloch vector distribution. This illustrates that the FWM contribution grows in time and that the local field induced phase inhomogeneity governs this process.

\subsubsection{Six-wave mixing quasi-Bloch vector}\label{sec:QBV_SWM}
Equipped with the quasi-Bloch vector picture we can retrieve the two most important SWM signal contributions by carrying out an analogue discussion. According to Fig.~\ref{fig:flow}, from all possible contributions after the third laser pulse, we only need the coherence $^{\phi_3}p_3^+$ and the occupations $^{|\Delta_{32}|}n_3^+$, $^{|\Delta_{32}-\Delta_{21}|}n_3^+$, and $^{|\Delta_{21}|}n_3^+$. Omitting all other terms we get
\begin{subequations}\begin{align}
p&\to i\frac\theta2 e^{i\phi_3} \,,\\
n&\to \Big[  \frac{\theta^2}{2} \cos(\Delta_{32}) 
		+\frac{\theta^2}2 \cos(\Delta_{21}) \notag\\
		&\qquad- \frac{\theta^4}8 V\tau \sin(\Delta_{32}-\Delta_{21})\label{eq:n_SWM}
		\Big]\,.
\end{align}\end{subequations}
To generate the SWM contribution $\sim \tau t$ we need to local-field mix the coherence filtered with respect to the SWM phase combination $2\phi_3-2\phi_2+\phi_1$ with the last term of the occupation in Eq.~\eqref{eq:n_SWM}, i.e., 
\begin{subequations}\begin{align}
^{\phi_3}p_3^+e^{-i\phi_{\rm SWM}}&= i \frac{\theta}2 e^{-i(\Delta_{32}-\Delta_{21})} \,,\\
^{|\Delta_{32}-\Delta_{21}|}n_3^+ &= - \frac{\theta^4}8 V\tau \sin(\Delta_{32}-\Delta_{21}) \,.
\end{align}\end{subequations}
From this we can directly read the respective initial quasi-Bloch vector
\begin{equation}\label{eq:QB_SWM_taut}
	\tilde{{\bm v}}_{{\rm SWM},\tau t}(t=0)
	= \frac{\theta}{2} \begin{pmatrix}
		\sin(\alpha)\\
		\sin(\alpha+\pi/2)\\
		-\frac14 \theta^3 V \tau \sin(\alpha)
	\end{pmatrix}\,,
\end{equation}

with $\alpha= \Delta_{32}-\Delta_{21}$. We see that this is the same Lissajous curve structure as in the previously discussed FWM case. The only differences are that the phase variation is now given by $\Delta_{32}-\Delta_{21}$, which still uniformly covers all angles, a tilt around the Im($\tilde{p}$)-axis, and a tilt angle such that the occupation distribution $\tilde{n}$ has an amplitude of $- \theta^4 V \tau/8$. The subsequent free dynamics are again governed by Eq. \eqref{eq:LF}.

Before discussing the shape and dynamics of this quasi-Bloch vector set we consider the other relevant SWM contribution $\sim t^2$. Therefore, we have a look at
\begin{subequations}\begin{align}
^{\phi_3}p_3^+e^{-i\phi_{\rm SWM}}&= i \frac{\theta}{2} e^{-i(\Delta_{32}-\Delta_{21})} \,,\\
^{|\Delta_{32}|}n_3^+ + ^{|\Delta_{21}|}n_3^+ &= \theta^2  \cos\left(\frac{\Delta_{32}-\Delta_{21}}{2}\right)\notag\\
&\quad\times \cos\left(\frac{\Delta_{32}+\Delta_{21}}{2}\right) \,.
\end{align}\end{subequations}

We are finally only interested in the coherence $^{\rm SWM}p_3$ describing the SWM signal. So, at this point we can already perform the integration
\begin{align}
&\int\limits_{-\pi}^\pi {\rm d}(\Delta_{32}+\Delta_{21}) \theta^2  \cos\left(\frac{\Delta_{32}-\Delta_{21}}{2}\right) \cos\left(\frac{\Delta_{32}+\Delta_{21}}{2}\right) \notag \\
&= 4 \theta^2  \cos\left(\frac{\Delta_{32}-\Delta_{21}}{2}\right) \,.
\end{align}
This leaves us with the quasi-Bloch vector for this SWM contribution reading
\begin{equation}\label{eq:QB_SWM_tt}
	\tilde{{\bm v}}_{{\rm SWM}, t^2}(t=0)
	= \frac{\theta}{2} \begin{pmatrix}
		\sin(\alpha)\\
		\sin(\alpha+\pi/2)\\
		8\theta \sin\left(\alpha/2+\pi/2\right)
	\end{pmatrix}\,,
\end{equation}
with $\alpha = \Delta_{32}-\Delta_{21}$, which is obviously a three dimensional Lissajous curve with the frequency ratio 2:2:1.

\begin{figure}[t]
	\centering
	\includegraphics[width=\columnwidth]{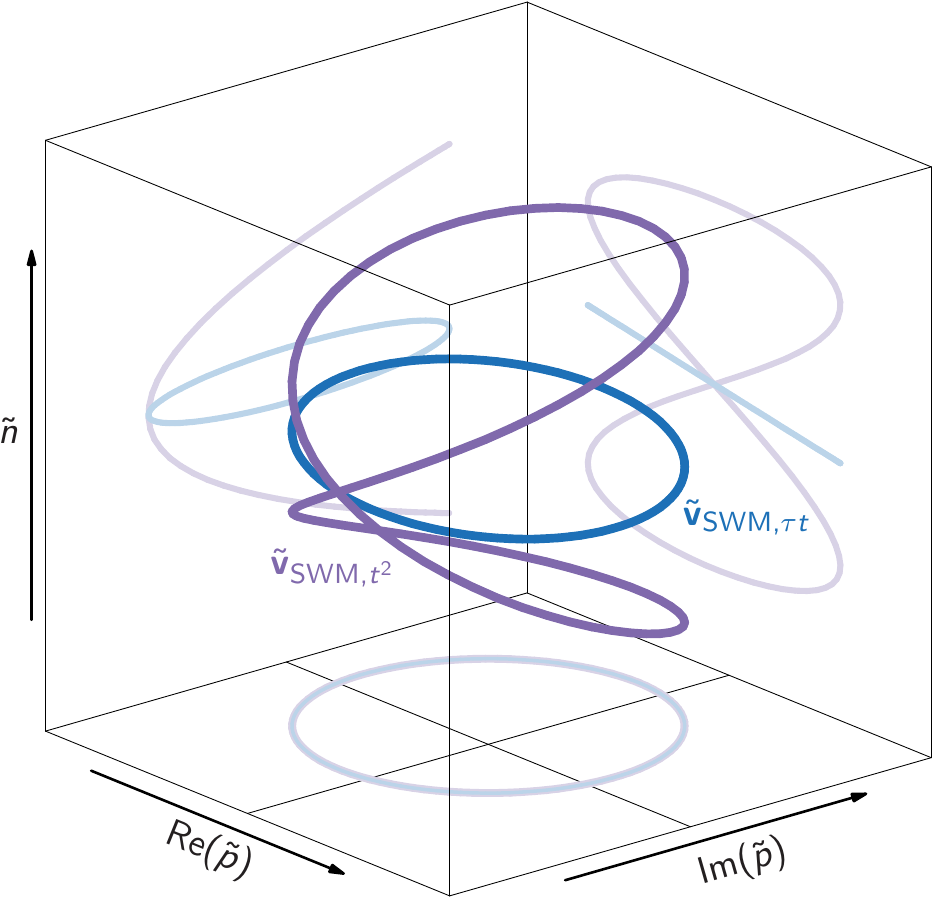}
	\caption{Quasi-Bloch vector distributions after the 3rd laser pulse. The $\sim t^2$ contribution (violet) and $\sim \tau t$ (blue) form Lissajous curves with frequency ratios 2:2:1 and 1:1:1, respectively.}\label{fig:Bloch_3D}
\end{figure}%

\begin{figure*}[h!]
	\centering
	\includegraphics[width=\textwidth]{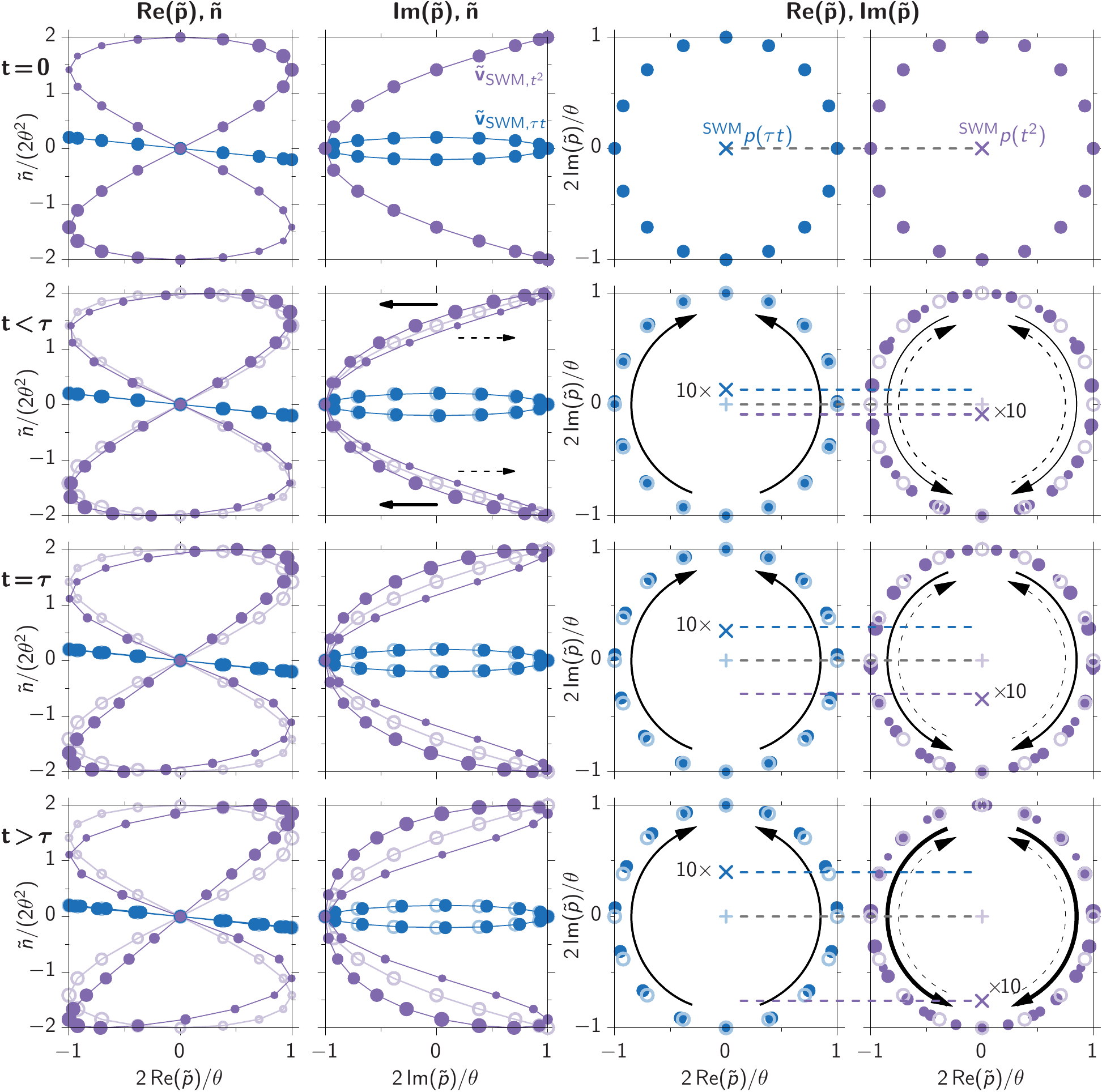}
	\caption{Projections of the quasi-Bloch vector representation of the destructive echo formation. Columns show projections of the quasi-Bloch vector defined in Eq.~\eqref{eq:Quasi_Bloch} as listed on the top, the rows show different real times as given on the left. The contribution $\sim t\tau$ is shown as blue, $\sim t^2$ as violet dots. The size of the violet dots represents the product $\tilde{n}\,{\rm Re}(\tilde{p})$, positive values are larger and negative ones smaller. The situation for $t=0$ is shows as bright circles for all other times. The black arrows indicate the movement of the quasi-Bloch vectors, solid (dashed) for positive (negative) $\tilde{n}\,{\rm Re}(\tilde{p})$. The final SWM coherences are marked as crosses in the two right columns.}\label{fig:Bloch}
\end{figure*}%

The two initial distributions of quasi-Bloch vectors given in Eqs.~\eqref{eq:QB_SWM_taut} and \eqref{eq:QB_SWM_tt} are depicted in Fig.~\ref{fig:Bloch_3D} in blue and violet, respectively. We vividly see the Lissajous curves leading to the distributions in $\tilde{n}$-direction. While the three dimensional perspective of the curves is appealing, we will consider the projections to the three different planes of the coordinate system when discussing the dynamics of the quasi-Bloch vectors, which are depicted in pale colors. These projections are shown in Fig.~\ref{fig:Bloch}, where different times are sorted in rows (increasing from top to bottom) and different perspectives in columns. 

First considering the $\sim \tau t$ contribution (blue) we find essentially the same situation as for the intermediate FWM contribution discussed in Fig.~\ref{fig:Bloch_FWM}. The only important difference is that for the SWM contribution the circle is tilted around the Im($\tilde{p}$)-axis. Following the same arguments as before the quasi-Bloch vector realizations move into the positive Im($\tilde{p}$)-direction. This movement is shown in the second column from the right in Fig.~\ref{fig:Bloch}. The initial homogeneous distribution is depicted by bright empty circles and for the considered times (written on the left) as filled dark circles. The movement is marked by the curved black arrow and the final SWM contribution after integration over all quasi-Bloch vector realizations as a dark blue cross. The speed of this movement is again constant and given by $V\tau$, which is proportional to the spread of $\tilde{n}$ and therefore given by the tilt angle of the circle.

For the $\sim t^2$ contribution (violet) the situation is more involved because the Lissajous curve is less trivial. In this situation the crossing point of the two loops lies at $\tilde{\bm v}_{{\rm SWM},t^2}=(0,{\rm Im}(\tilde{p}),0)$. Because of the vanishing occupation this point will not move in the following propagation and the Im$(\tilde{p})$-axis is again the symmetry axis of the dynamics. Consequently, the opposing Re$(\tilde{p})$ values compensate each other, such that the final SWM contribution is again purely imaginary. Therefore, we have to consider the movement into the direction governed by the product $\tilde{n}\,{\rm Re}(\tilde{p})$. When looking at the left column in Fig.~\ref{fig:Bloch} the distribution reaches values with $\tilde{n}\,{\rm Re}(\tilde{p})>0$ (large dots) and $\tilde{n}\,{\rm Re}(\tilde{p})<0$ (small dots). Following Eq.~\eqref{eq:LF} the large dots move into the negative Im($\tilde{p}$)-direction while the small ones move into the positive Im($\tilde{p}$)-direction as indicated by the black horizontal arrows in the Im($\tilde{p}$),$\tilde n$-plane at $t<\tau$. In the complex $\tilde{p}$-plane this leads to two counter movements marked as arrows in the right column. The small dots move up (dashed curved arrows) and the large dots move down (solid curved arrows). Initially both of the these movements start with the same velocity because large and small dots have the same absolute values $|\tilde{n}\,{\rm Re}(\tilde{p})|$, i.e., the Lissajous curve is symmetric. Therefore, the integrated SWM signal starts with a vanishing velocity for small $t\approx 0$. However, during the following dynamics the quasi-Bloch vectors with positive Im($\tilde{p}$)-velocity (small dots) rotate collectively in such a way that this part of the curve obtains smaller occupations. This can be seen by focussing on the bottom-left and upper-right corners of the panels in the left column. Accordingly, those with negative Im($\tilde{p}$)-velocity (large dots) obtain larger $|\tilde{n}|$ as can be seen in the second column [Im($\tilde{p}$),$\tilde{n}$-plane] marked by the horizontal black arrows. In the left Lissajous curve this means that the large dots move towards the top left and bottom right corner of the plot, while the small dots move inwards. Consequently, the negative Im($\tilde{p}$)-velocity component gains speed because $|\tilde{n}\,{\rm Re}(\tilde{p})>0|$ (large dots) grows, while the positive component slows down as $|\tilde{n}\,{\rm Re}(\tilde{p})<0|$ (small dots) shrinks. In the complex $\tilde{p}$-plane (right column) this means that the upwards movement of the quasi-Bloch vectors represented by small dots (dashed arrows) slows down for increasing times $t$ while the downwards movement of the large dots (solid arrows) speeds up. In summary, for the integrated SWM coherence, marked by the violet cross, we find an acceleration towards negative Im($\tilde{p}$)-values given by $-V^2t^2$.

\begin{table*}[h]
\centering
\caption{\label{tab:compare}
Comparison of criteria defining the photon echo in FWM and the destructive photon echo in SWM.
}
\begin{tabular}{c|c|c}
 & echo & destructive echo\\
 & (FWM) & (SWM) \\
\hline
\multirow{2}{*}{timing} & delay + internal & delay + internal \\
	& dynamics & dynamics\\
\hline
\multirow{2}{*}{interference} & constructive & destructive \\
	& (1 path) & (2 paths)\\
\hline
\multirow{2}{*}{source} & inhomogeneous & local field induced\\
	& broadening & phase inhomogeneity
\end{tabular}
\end{table*}

Exactly at $t=\tau$ in the third row the two integrated SWM contributions marked in the third and forth column have the same distance from the center as indicated by the dashed horizontal lines which show that they compensate each other. At this time the destructive echo appears due to the destructive quantum interference of all depicted quasi-Bloch vectors. For even larger times $t<\tau$ (bottom row) the accelerated $\sim t^2$ contribution is stronger than the one with constant velocity $\sim \tau t$ and the SWM is non-vanishing again.

\section{Conclusion}

To conclude we again compare the traditional photon echo in FWM with the newly discovered {\it destructive photon echo} in SWM. As summarized in Tab.~\ref{tab:compare} we define three criteria that characterize the echo formation. Firstly, the timing of the photon echo formation is basically determined by the chosen delay $\tau$ between the two laser pulses. Slight deviations from the exact $t=\tau$ timing are used to learn about internal dynamics of the studied system~\cite{poltavtsev2016photon}. The same holds for the destructive echo in SWM. As discussed in detail the two main contributions compensate each other exactly at $t=\tau$. Nonetheless, in the detected and simulated full signal dynamics in Fig.~\ref{fig:antiecho} the depression happens at $t>\tau$. The reason is that exciton decay dynamically changes the slope of the destructive echo. Secondly, the photon echo appears because all considered Bloch vectors form a constructive interference. To explain this effect only the conventional FWM signal has to be considered, which would appear as a single path in a flow chart like the one in Fig.~\ref{fig:flow} (see also Ref.~\cite{hahn2021infl}). In the case of the destructive echo we have shown that we need two paths which interfere destructively to explain the novel echo effect. Finally, the fundamental source of the photon echo is any sort of inhomogeneity, which might appear in space by a locally varying exciton energy~\cite{koch1993subpicosecond}, or in time via external noise that acts on the optical transition energy~\cite{langbein2010coherent}. In the case of the destructive echo we have shown that the local field together with a variation of the applied laser pulse phases results in a spectral broadening. As discussed in Ref.~\cite{hahn2021infl} this local field induced phase inhomogeneity can already be detected in FWM spectra. This summary shows that the criteria defining the traditional photon echo can also be applied to the destructive echo effect. Therefore, we conclude that we really found a new photon echo effect.

In addition we have developed a quasi-Bloch vector picture to illustrate the generation of the destructive echo inspired by the instructive Bloch vector image of the photon echo. We found that the SWM-relevant quasi-Bloch vectors are distributed along Lissajous curves which adds surprising aesthetics to the involved Bloch vector dynamics in SWM.

Following this proof-of-principle demonstration of a new photon echo effect the natural next tasks will be to further explore the destructive photon echo's application possibilities.\linebreak One obvious step will be to measure photon echos at various spots of the sample and thereby get a measure for the spatial distribution of the local field coupling strength. We know that different pulse sequences in FWM probe different observables~\cite{mermillod2016dynamics}. Therefore we hope that novel pulse sequences in higher wave-mixing spectroscopy will be designed that should allow to access parameters governing the nonlinear light-\linebreak induced dynamics in semiconductors, like the strength of interaction mechanisms, that are not easily accessible with traditional techniques. These developments are particularly attractive for materials with a large nonlinearity promoting strong wave mixing responses like TMDCs.


\section{Experimental Section}
In the experiment, we employ a mode-locked laser ({\it Tsunami Femto} provided by {\it Spectra Physics}) generating femtoscond \linebreak pulses at a repetition rate of 76~MHz. The wavelength is tuned to 752~nm, which corresponds to the A exciton transition in the investigated hBN/MoSe$_2$/hBN heterostructure at $T=5$~K. The sample is kept in an optical He-flow cryostat and microspectroscopy is performed using an {\it Olympus} microscope objective ({\it LCPLN50XIR}, numerical aperture of 0.65), which is installed on an XYZ piezo stage from {\it Physik Instrumente}. To perform the multi-pulse, heterodyne experiments, the initial pulse train from the laser source is split into three and each of these beams is focussed into a separate acousto-optic modulator (AOM). The AOMs are driven at distinct radio-frequencies of $\Omega_1=79$, $\Omega_2=80.77$, and $\Omega_3=81$~MHz, such that the deflected beams acquire corresponding frequency upshifts. Next, the time delays between the first two ($\tau_{12}$) and the last two beams ($\tau_{23}$) are introduced by a pair of mechanical delay stages. In addition, a grating-based pulse shaper is employed to correct the temporal chirp of the pulses, when passing through optical elements of the setup, especially AOMs and the microscope objective. The such prepared pulse sequence is then recombined into the same spatial mode and focussed at the heterostructure sample reaching a diffraction limited spot diameter of around 0.8~\textmu m. The nonlinear optical signal from the sample is retrieved in the back-reflectance geometry. To isolate the desired wave-mixing response another AOM is used, operating at the heterodyne frequency generated by a home-made analogue radio-frequency mixing electronics, assembled from individual components provided by {\it Mini Circuits}. For example, the studied SWM signal is detected at ($\Omega_{\rm SWM}=2\Omega_3 - 2\Omega_2+\Omega_1$)=81.54~MHz. After being deflected from the AOM the signal is frequency downshifted by $\Omega_{\rm SWM}$ and the unique wave-mixing component  under consideration carries the original frequency from the laser source. The signal can now interfere with a reference beam from the same laser source, which propagates in the vicinity of the driving beams. At the same time all other optical response components present in the signal reflected from the sample are still modulated in the MHz range and thus average out completely during a single acquisition time of a few ms. Importantly, the mixing AOM operates in a Bragg configuration. Hence, simultaneously the reference beam gets deflected onto the signal, receiving a frequency upshift by $\Omega_{\rm SWM}$. This again allows for interference with the signal, which was not deflected by the AOM (operating at $\Omega_{\rm SWM}$). In total we generate two beams in which the considered wave-mixing signal is interfering with the reference pulse, but with the opposite phase, as required by energy conservation. The spectrally-resolved interference is obtained with an imaging spectrometer ({\it Princeton Instruments} with a focal length of 750~mm) and detected with an CCD camera ({\it PIXIS} from {\it Princeton Instruments}, with an {\it eXcelon} coating). A background free, shot-noise limited detection is achieved by a balanced detection and multi-acquisition. Namely, we exploit the Bragg configuration and subtract the two mixed beams impinging at different positions on the CCD. In addition, the phase of the mixing AOM is cycled between 0 and $\pi$ to overcome any classical noise from the CCD camera. The wave-mixing signal amplitude and phase are obtained from the interferogram by performing spectral interferometry.



\section*{Acknowledgements}  
T.H. thanks the German Academic Exchange Service (DAAD) for financial support (No. 57504619).
D.W. thanks the Polish National Agency for Academic Exchange (NAWA) for financial support within the ULAM program\linebreak (No. PPN/ULM/2019/1/00064).
T.H., D.W., T.K., and P.M. acknowledge support from NAWA under an APM grant.
M.P, D.V., and M.B thank the EU Graphene Flagship project.
M.P, D.V, K.N, and J.K acknowledge support from CNRS via IRP “2D materials”.
M.P. and  K.N  acknowledge support from TEAM programme of the Foundation for Polish Science, co-financed by the EU within the ERDFund.
M.P. is supported by the ESF under the project CZ.02.2.69/0.0/0.0/20\_079/0017436.
 K.W. and T.T. acknowledge support from the Elemental Strategy Initiative conducted by the MEXT, Japan (Grant Number\linebreak JPMXP0112101001) and JSPS KAKENHI (Grant Numbers\linebreak 19H05790 and JP20H00354).


%
\section*{References}

\end{document}


\maketitle

\begin{affiliations}\small
 \item Institut of Solid State Theory, University of M\"unster,\\ 48149 M\"unster, Germany
 \item Department of Theoretical Physics, Wroc\l{}aw University of Science and Technology,\\ 50-370~Wroc\l{}aw, Poland
 \item Laboratoire National des Champs Magn\'{e}tiques Intenses, CNRS-UGA-UPS-INSA-EMFL,\\ 38042 Grenoble, France
 \item Central European Institute of Technology, Brno University of Technology,\\ 61200 Brno, Czech Republic
 \item Institute of Experimental Physics, Faculty of Physics, University of Warsaw,\\ 02-093 Warszawa, Poland
 \item Research Center for Functional Materials, National Institute for Materials Science,\\ Tsukuba 305-0044, Japan
 \item International Center for Materials Nanoarchitectonics, National Institute for Materials Science,\\ Tsukuba 305-0044, Japan
 \item Universit\'{e} Grenoble Alpes, CNRS, Grenoble INP, Institut N\'{e}el,\\ 38000 Grenoble, France
 \item[$^\ast$] daniel.wigger@pwr.edu.pl
\end{affiliations}

\begin{abstract}\small
This document includes following content:
\begin{itemize}
\item[S1] All SWM contributions in the $\boldsymbol{\chi^{(5)}}$-regime
\item[S2] Impact of the local field strength
\item[S3] Impact of excitation induced dephasing
\item[S4] Impact of intervalley scattering
\item[S5] Impact of a non-vanishing delay $\boldsymbol{\tau_{12}}$
\item[S6] Impact of inhomogeneous broadening
\end{itemize}
\end{abstract}
\thispagestyle{empty}

\twocolumn
\footnotesize


\section{All SWM contributions in the $\boldsymbol{\chi^{(5)}}$-regime}\label{sec:A_equations}
To find all phase combinations which lead to the SWM-phase, we go back to the transformation rules in Eqs.~(2):
\begin{subequations}\label{eq:pulse_appendix}\begin{align}
	p^+ &= p^- \cos^2\left(\frac{\theta}{2}\right) + \frac{i}{2}\sin(\theta)(1-2n^-) e^{i\phi} \notag\\
		&\qquad\qquad + \sin^2\left(\frac{\theta}{2}\right) p^{-*} e^{i2\phi}\notag\\
		&\approx p^-\left(1-\frac{\theta^2}{4}\right) + i \frac{\theta}{2} (1-2n^-) e^{i\phi} + \frac{\theta^2}{4}p^{-*}e^{i2\phi} \, ,\\[2mm]
	n^+ &= n^- + \sin^2\left(\frac{\theta}{2}\right) (1-2n^-) +\sin(\theta){\rm Im}\left(p^-e^{-i\phi}\right) \notag\\
		&\approx n^- + \frac{\theta^2}{4}(1-2n^-) + \theta{\rm Im}\left(p^-e^{-i\phi}\right) \, ,
\end{align}\end{subequations}
where we also included second order contributions in the optical field, which will become relevant later. We again set the pulse areas of all three pulses to be equal $\theta_1= \theta_2=\theta_3=\theta$. After the first pulse, up to $\mathcal O(\theta^2)$ we get
\begin{subequations}\begin{align}
	p_1^+ &= i\frac \theta2  e^{i\phi_1} \,,\\
	n_1^+ &= \frac {\theta^2}2 \,.
\end{align}\end{subequations}
After the second pulse the state reads up to $\mathcal O(\theta^3)$
\begin{subequations}\begin{align}
	p_2^+ &= i\frac \theta2 \left(e^{i\phi_1} + e^{i\phi_2}\right) - i\frac {\theta^3}8 \left[e^{i\phi_1}  + 2e^{i\phi_2} + e^{i(2\phi_2-\phi_1)}\right]\,,\\
	n_2^+ &= \frac {\theta^2}2 [1+\cos(\phi_2-\phi_1)]\,.
\end{align}\end{subequations}
Anticipating the transformation of the third pulse, we see that from the third-order polarization ($\sim\theta^3$) only the part with the FWM-phase $2\phi_2-\phi_1$ needs to be taken into account. Thus, collecting all the linear polarizations and the FWM polarization we obtain with Eq.~(5) the time evolution during the delay $\tau$
\begin{subequations}\begin{align}
	p_3^- &= \left[ i\frac \theta2 \left(e^{i\phi_1} + e^{i\phi_2}\right) - i\frac {\theta^3}8 e^{i(2\phi_2-\phi_1)} \right]e^{-\beta\tau}\notag\\
	&\quad  -iV\tau \left[ i\frac \theta2 \left(e^{i\phi_1} + e^{i\phi_2}\right)\right] \frac{\theta^2}2 \cos(\phi_2-\phi_1) e^{-\beta\tau} \,,\notag\\
	&\to \left[ i\frac \theta2 \left(e^{i\phi_1} + e^{i\phi_2}\right) - i\frac {\theta^3}8 e^{i(2\phi_2-\phi_1)} \right]e^{-\beta\tau} \notag\\
	&\qquad\qquad+ V\tau \frac{\theta^3}8 e^{i(2\phi_2-\phi_1)} e^{-\beta\tau} \,,\\
	n_3^- &= \frac {\theta^2}2 [1+\cos(\phi_2-\phi_1)]\,.
\end{align}\end{subequations}
%
\begin{figure*}[ht]
	\centering
	\includegraphics[width=\textwidth]{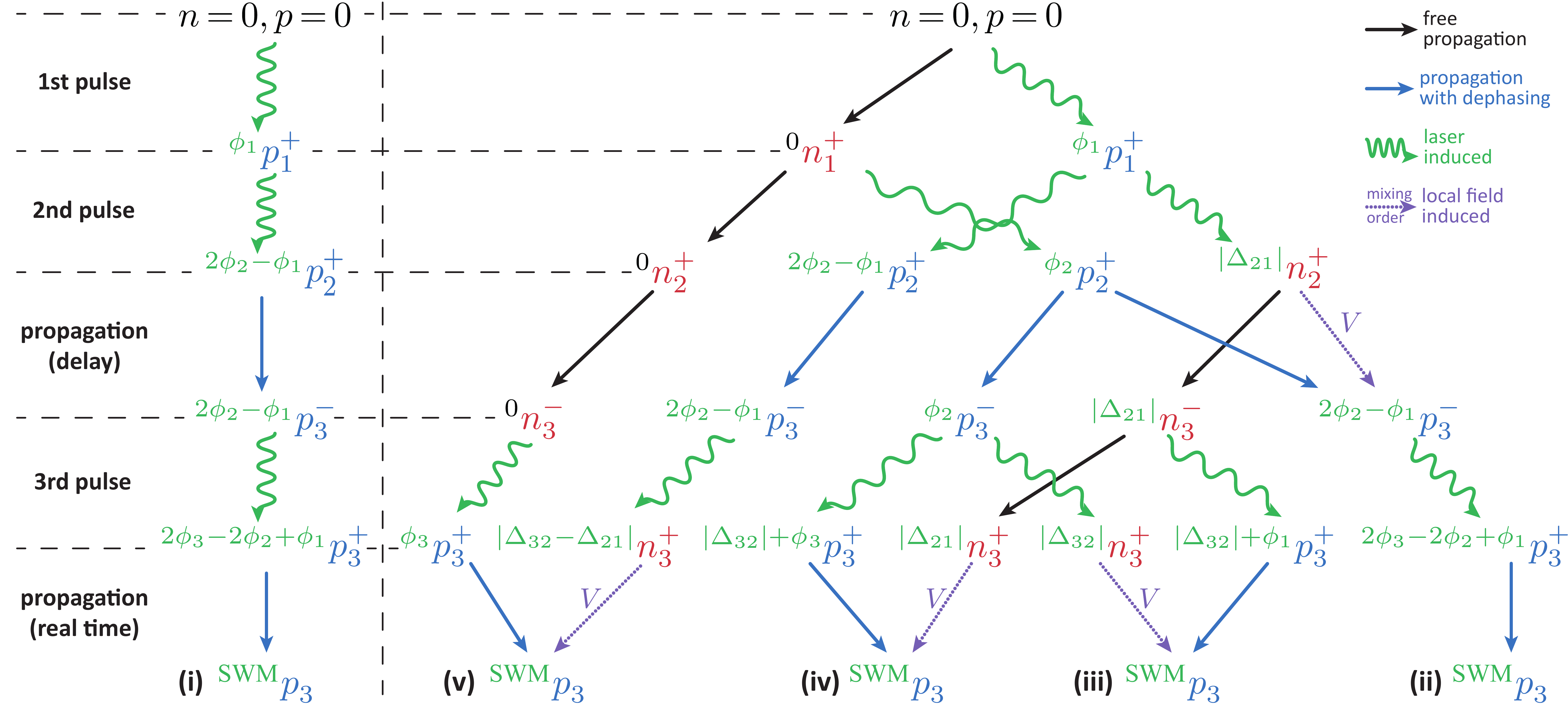}
	\caption{Flow charts for the SWM signal contributions without local field coupling (left) and proportional to $V$ (right). Same as Fig.~5 in the main text. The labels (i) -- (v) refer to the respective contributions in Eqs.~(\ref{eq:SWM_contr}).}\label{fig:flow_1}
\end{figure*}%
Again, we have omitted all terms which will not contribute to the SWM signal in the $\chi^{(5)}$-regime. Note, that due to intervalley scattering, the occupation $n_3^-$ is split between the valleys as discussed in the main text. The third pulse then transforms polarization and occupation into

\begin{subequations}\begin{align}
	p_3^+ &= i\frac \theta2  \left( e^{i\phi_1}e^{-\beta\tau} +  e^{i\phi_2}e^{-\beta\tau} + e^{i\phi_3} \right)\notag\\
		& - i\frac{\theta^3}4 e^{i(\phi_3-\phi_2+\phi_1)} \notag\\
		& -  \frac{\theta^3}8 e^{i(2\phi_2-\phi_1)} \left( i - V\tau\right)e^{-\beta\tau} \notag\\ 
		& - i\frac{\theta^3}8 \left[ e^{i(2\phi_3-\phi_1)} + e^{i(2\phi_3-\phi_2)} \right]e^{-\beta\tau} \notag\\
		&+ \frac{\theta^5}{32} e^{i(2\phi_3 - 2\phi_2 +\phi_1)} \left(i + V\tau\right)e^{-\beta\tau} \,,\label{eq:p3}\\
	 n_3^+ &= \frac {\theta^2}2 \big[\cos(\phi_2-\phi_1)  + \cos(\phi_3-\phi_1)e^{-\beta \tau} \notag\\
	 	&\qquad\qquad+\cos(\phi_3-\phi_2)e^{-\beta \tau}\big]\notag\\
		&-\frac{\theta^4}8 \cos(\phi_3-2\phi_2+\phi_1)e^{-\beta\tau} \notag \\
		&- \frac{\theta^4}8  V\tau \sin(\phi_3-2\phi_2+\phi_1)e^{-\beta\tau}\,.
\end{align}\end{subequations}

Note, that we have again omitted all terms that are irrelevant for the final SWM signal, especially constant terms for the occupation that do not carry any phase information. Next, we consider the time-evolution after the third pulse, where we can already sort for the SWM phase $2\phi_3-2\phi_2+\phi_1$. In Fig.~5 in the main text we schematically depict the origin of the contributions with two local-field mixing processes for $ \mathcal O(V^2)$. We disentangle the different contributions with $\mathcal O(V^1)$ and the SWM polarization from the pure TLS without local-field mixing with $\mathcal O(V^0)$ by the help of the flow charts in Fig.~\ref{fig:flow_1}.

Starting with the polarization in \eqref{eq:p3} that already carries the SWM phase, their signal contribution reads 
\begin{subequations}\label{eq:SWM_contr}\begin{align}
	^\text{SWM}p/\theta^5 = \frac i{32} e^{-\beta\tau} + \frac 1{32} V\tau e^{-\beta\tau}\,.
\end{align}
The first part represents the SWM signal from the pure TLS as it is independent of $V$. Its origin is depicted in Fig.~\ref{fig:flow_1} on the left side (i) and we directly see that the coherence is detected when scanning the delay between the second and third pulse as mentioned in the main text. The second term (ii) is proportional to $V\tau$, therefore it stems from the local-field mixing between $^{\phi_2}p_2^+$ and $^{|\Delta_{21}|}n_2^+$ during the delay propagation, depicted is the most right path in Fig.~\ref{fig:flow_1}.

From the FWM-polarizations created by the third pulse, we get the two SWM contributions (iii) and (iv) by a single local field mixing process in the real time propagation. According to Eq.~(5) we have to calculate $p(t)=-iVt p_0 n_0$ for single and $p(t) = -(Vt)^2p_0n_0^2/2$ for double mixing. (iii) is created by mixing the three-pulse FWM term $^{\Delta_{32} + \phi_1}p_3^+ = {}^{\phi_3 - \phi_2 + \phi_1}p_3^+$ and $^{|\Delta_{32}|}n_3^+$ resulting in
\begin{align}
	^\text{SWM}p/\theta^5 &= -iVt \left(-\frac i4 \right)\left(\frac14 e^{-\beta\tau}\right) \notag\\
	&= -\frac 1{16}Vt e^{-\beta\tau}\,.
\end{align}
The second one (iv) is created by the two-pulse FWM term $^{2\phi_3 - \phi_2}p_3^+$ mixed with $^{|\Delta_{21}|}n_3^+$ leading to
\begin{align}
	^{\rm SWM}p/\theta^5 &= -iVt \left(-\frac i8e^{-\beta\tau}\right)\left( \frac14\right) \notag\\
	& = -\frac 1{32} Vt e^{-\beta\tau}\,.
\end{align}

Finally, contribution (v) is created by local field mixing $^{\phi_3}p_3^+$ with $^{|\Delta_{23}-\Delta_{12}|}n_3^ +$ resulting in
\begin{align}
	^{\rm SWM}p/\theta^5 &= -iVt \left(\frac i2\right) \left( -\frac 1{16} e^{-\beta\tau}\right) \notag\\
	&= -\frac 1{32} Vt e^{-\beta\tau} \,.
\end{align}\end{subequations}

For completeness we also give the three SWM contributions with $\mathcal O(V^2)$ here. As illustrated in Fig.~5 in the main text they are retrieved by double local field mixing between $^{\phi_1}p_3^+$ and $^{\Delta_{32}}n_3^+$
\begin{subequations}\begin{align}
	^{\rm SWM}p/\theta^5 &= -\frac12(Vt)^2 \left(\frac i2 e^{-\beta\tau} \right)\left( \frac 1{4} e^{-\beta\tau}\right)^2 \notag\\
	&= -\frac i {64} (Vt)^2 e^{-3\beta\tau}\,,
\end{align}
between $^{\phi_3}p_3^+$, $^{\Delta_{32}}n_3^+$, and $^{\Delta_{21}}n_3^+$
\begin{align}
	^{\rm SWM}p/\theta^5 &=  -\frac12(Vt)^2 \left(\frac i2\right) 2 \left( \frac 1{4} e^{-\beta\tau} \right)\left( \frac 1{4} \right) \notag\\
	&= -\frac i{32} (Vt)^2 e^{-\beta\tau}\,,
\end{align}
where the factor 2 stems from the sum of the two occupations, and finally the single local field mixing between $^{\phi_3}p_3^+$ and $^{\Delta_{32}-\Delta_{21}}n_3^+$
\begin{align}
	^{\rm SWM}p/\theta^5 &= -iVt \left(\frac i2\right)\left( -\frac 1{16 i} V\tau e^{-\beta\tau} \right) \notag\\
	&= \frac i{32} V^2 t\tau e^{-\beta\tau}\,.
\end{align}\end{subequations}

Collecting all the expressions finally yields
\begin{align}\label{eq:swm_analyt_2}
	 ^{\rm SWM}p &= \left(\frac{\theta}{2}\right)^5 \bigg[ i + V( \tau - 4t ) \notag\\
	 	&\qquad - \frac i2(Vt)^2 e^{-2\beta\tau} + i V^2t(\tau-t)  \bigg] e^{-\beta(t+\tau)}
\end{align}
for the SWM signal in the $\chi^{(5)}$-regime.

\begin{figure}[t]
	\centering
	\includegraphics[width=\columnwidth]{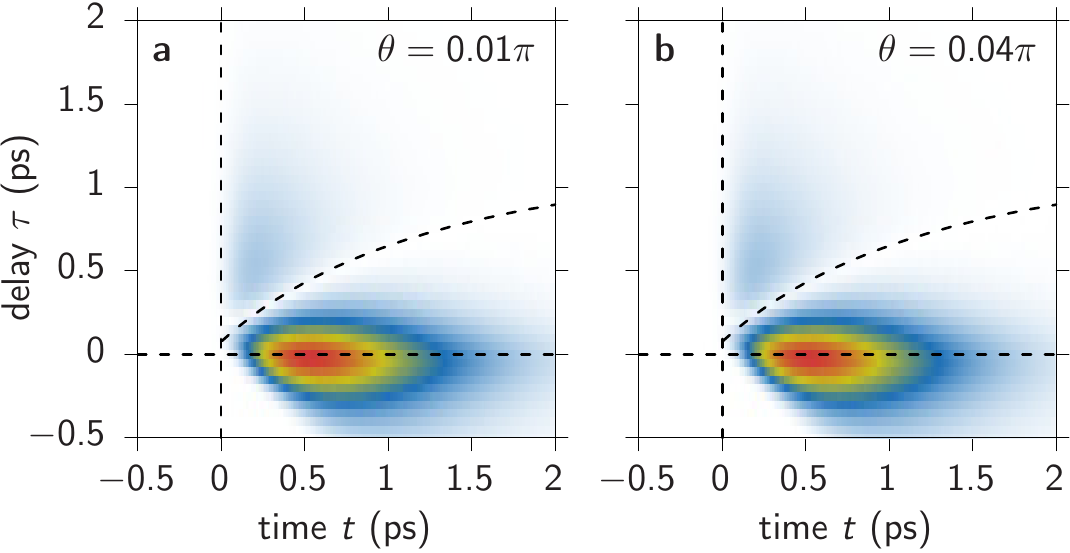}
	\caption{SWM dynamics for different pulse areas $\theta$ being halved (a) or doubled (b) compared to Fig.~4(b) in the main text.}\label{fig:SWM_theta}
\end{figure}%

From Eq.~(7)/(\ref{eq:swm_analyt_2}) we see that in the lowest order of the light-matter coupling, the signal amplitude scales with the fifth power of the pulse area. Therefore, operating in the low excitation regime the SWM dynamics do not depend on the choice of $\theta$. To confirm this in Fig.~\ref{fig:SWM_theta} we plot the same SWM dynamics as in Fig.~4(b) in the main text but with halved (a) and doubled (b) pulse area. There are no obvious differences between the two simulations. This confirms that the chosen pulse area does not affect the signal dynamics and that the determined local field coupling is a reasonable quantity to characterize the dynamics.

\section{Impact of the local field strength}\label{sec:A_V}
\begin{figure}[h]
	\centering
	\includegraphics[width=\columnwidth]{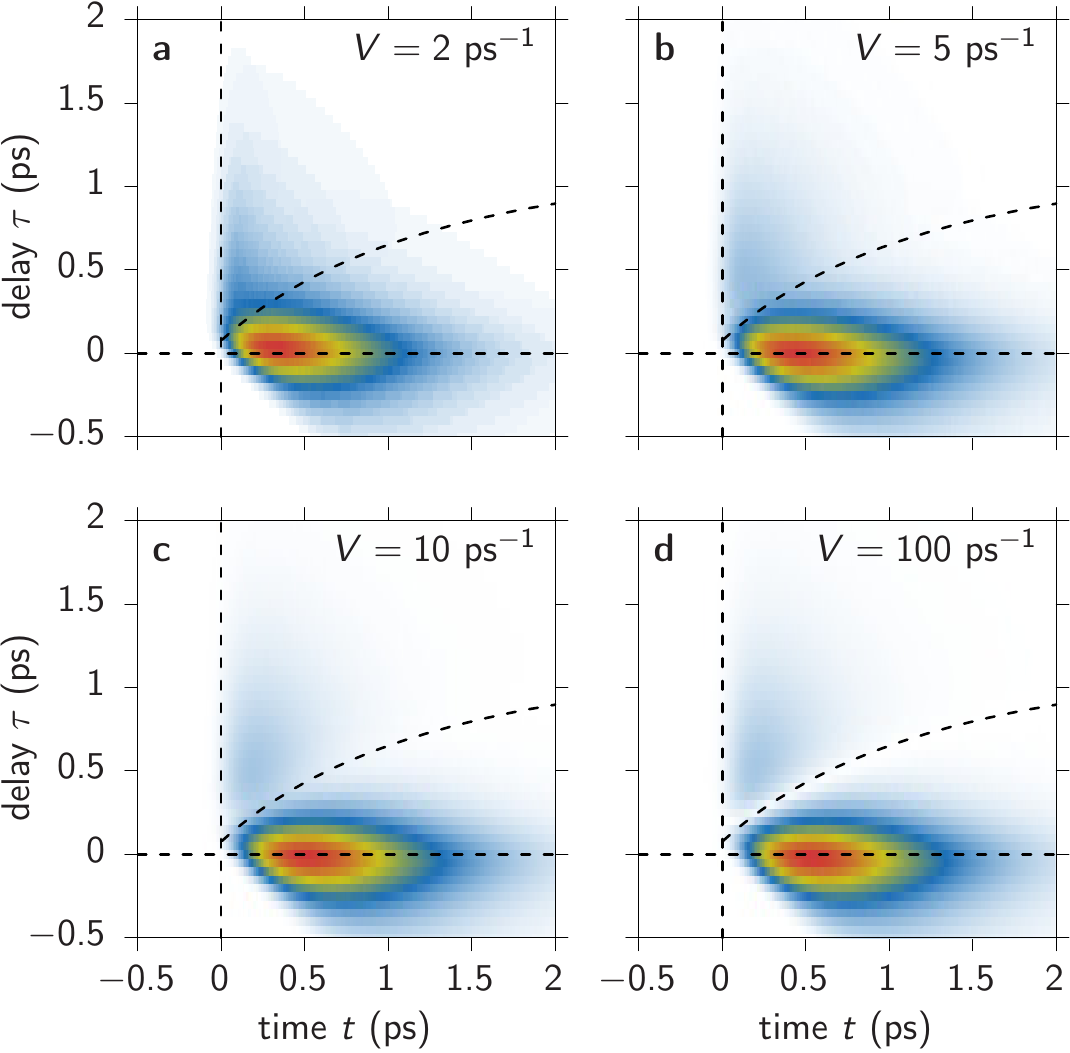}
	\caption{Comparison of the SWM dynamics for different local field strengths $V$ increasing from (a) to (d), where (d) is the example from Fig.~4(b) in the main text.}\label{fig:SWM_V}
\end{figure}
From the analytical solution, we find the destructive echo only in the highest possible order, i.e., $V^2$. To understand, how the local field strength influences the destructive echo, in Fig.~\ref{fig:SWM_V} we show the SWM dynamics for different values of $V$ increasing from (a) to (d). While for $V=2\,$ps$^{-1}$~(a) and $V=5\,$ps$^{-1}$~(b) no destructive echo is visible, for $V=10\,$ps$^{-1}$ (c), the depression is already visible. For comparison Fig.~\ref{fig:SWM_V}(d) shows the case with $V=100$~ps$^{-1}$ from the main text. This again confirms, that the $V^2$-order has to dominate the signal to observe the destructive echo. For $V=10$~ps$^{-1}$ in Fig.~\ref{fig:SWM_V}(c) we see that the destructive echo follows the dynamics of the minimum approximated by Eq.~(9) shown as dashed curved line. This confirms our finding that this behavior is not affected by the strength of the local field coupling $V$.

\section{Impact of excitation induced dephasing}\label{sec:A_EID}
\begin{figure}[h]
	\centering
	\includegraphics[width=\columnwidth]{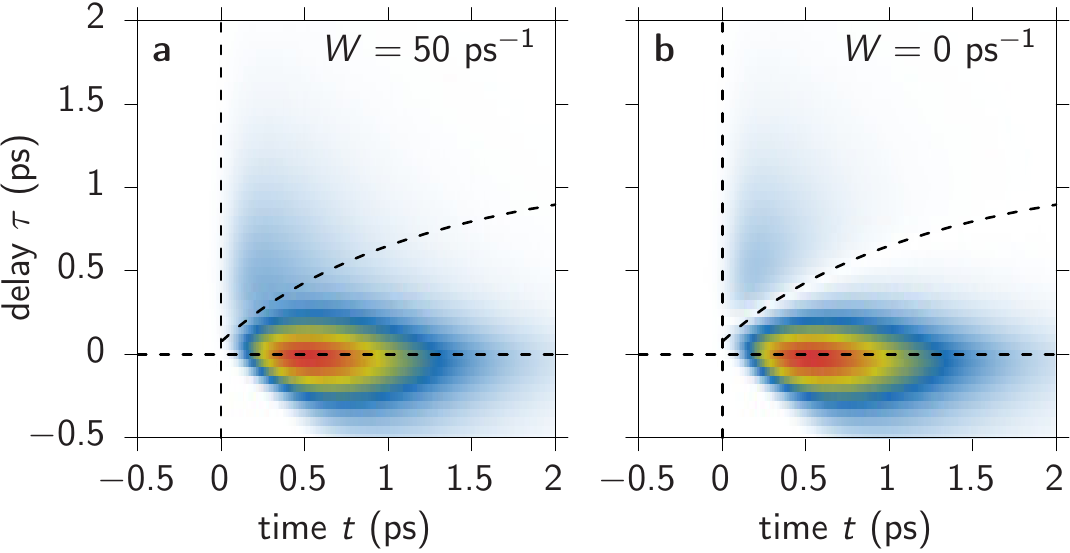}
	\caption{Impact of a non-vanishing excitation induced dephasing with $W=50$~ps$^{-1}$ (a) and (b) is the example from Fig.~4(b) in the main text with $W=0$.}\label{fig:SWM_W}
\end{figure}
In addition to the local field coupling which shifts the transition energy depending on the exciton occupation, we can also take excitation induced dephasing (EID) into account~\cite{rodek2021local}, where the dephasing grows with the occupation $\beta_\text{total} = \beta + W(n+n')$. For an EID on the same order of magnitude as the local field the destructive echo effect is strongly suppressed as illustrated in Fig.~\ref{fig:SWM_W}(a) for $W=50\,$ps$^{-1}$. As reference in Fig.~\ref{fig:SWM_W}(b) we show the case without EID ($W=0$) from the main text [Fig.~4(b)]. Obviously, the EID prohibits the development of a destructive photon echo.
\begin{figure}[h]
	\centering
	\includegraphics[width=0.85\columnwidth]{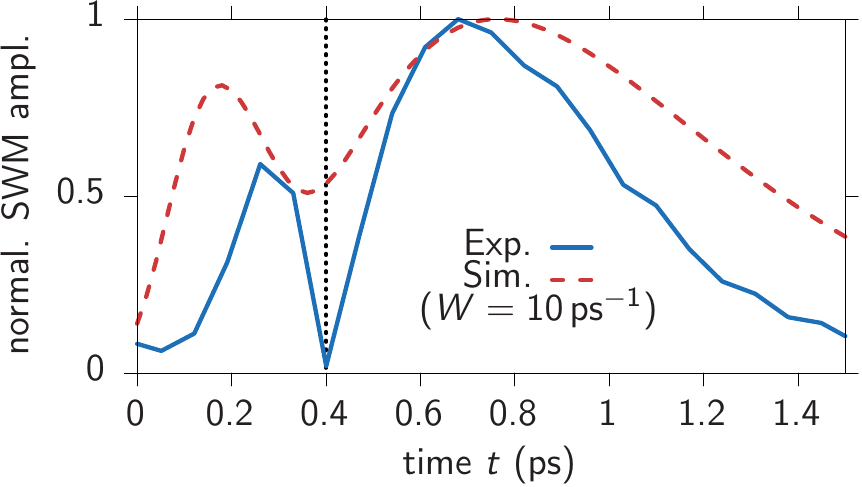}
	\caption{Impact of a non-vanishing excitation induced dephasing with $W=10$~ps$^{-1}$ on the SWM amplitude dynamics for $\tau=0.35$~ps. Experiment as solid blue line and simulation as dashed red line. The measurement is the same as in Fig.~7(a).}\label{fig:SWM_dyn_W}
\end{figure}%

In Fig.~\ref{fig:SWM_dyn_W} we plot the same SWM amplitude dynamics as in Fig.~7(a) in the main text but in the simulation we consider a EID of $W=10$~ps$^{-1}$, which is much smaller than the applied local field coupling of $V=100$~ps$^{-1}$. We see that the destructive photon echo minimum is significantly less pronounced when taking the EID into account. Therefore, to maintain the good agreement between theory and experiment, we have excluded EID from the discussion. We can conclude that the EID should not have a significant impact in our system.\\

\section{Impact of intervalley scattering}\label{sec:A_lambda}
\begin{figure}[h]
	\centering
	\includegraphics[width=\columnwidth]{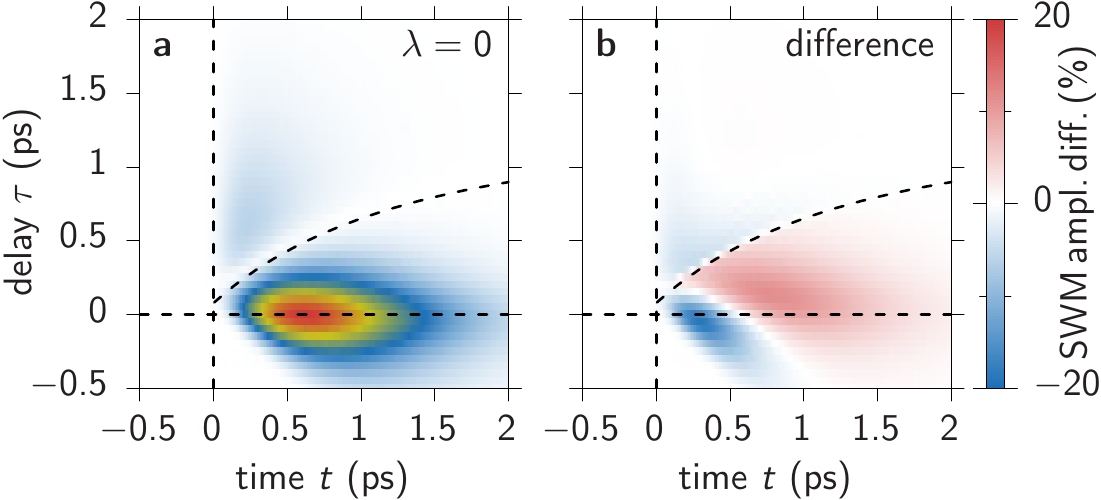}
	\caption{Impact of a vanishing intervalley scattering $\lambda~=~0$ on the SWM amplitude dynamics. (a) SWM amplitude dynamics. (b) Difference of the SWM amplitude with respect to the case with $\lambda=4$~ps$^{-1}$ from Fig.~4(b) in the main text.}\label{fig:SWM_lambda}
\end{figure}

In the context of the approximation in Eq.~(4b) we estimated that the intervalley scattering should only have an impact for small delays. To confirm this we calculate the SWM amplitude dynamics for $\lambda=0$ as depicted in Fig.~\ref{fig:SWM_lambda}(a). Compared to Fig.~4(b) in the main text we cannot find significant deviations. To highlight the changes when setting $\lambda=0$ in Fig.~\ref{fig:SWM_lambda}(b) we plot the difference with respect to the case with $\lambda=4$~ps$^{-1}$. We indeed find, that the deviations are on the order of only $\approx10\%$ and mainly restricted to delays $|\tau|<0.5$~ps. Interestingly, we find that the amplitude difference has a change of sign that exactly follows the approximated evolution of the destructive echo minimum marked by the dashed line [see Eq.~(9) in the main text]. Note, that the depicted color maps were calculated numerically employing the full model and taking a non-vanishing pulse duration into account.

\section{Impact of a non-vanishing delay $\boldsymbol{\tau_{12}}$}\label{sec:tau12a}
\begin{figure}[h]
	\centering
	\includegraphics[width=\columnwidth]{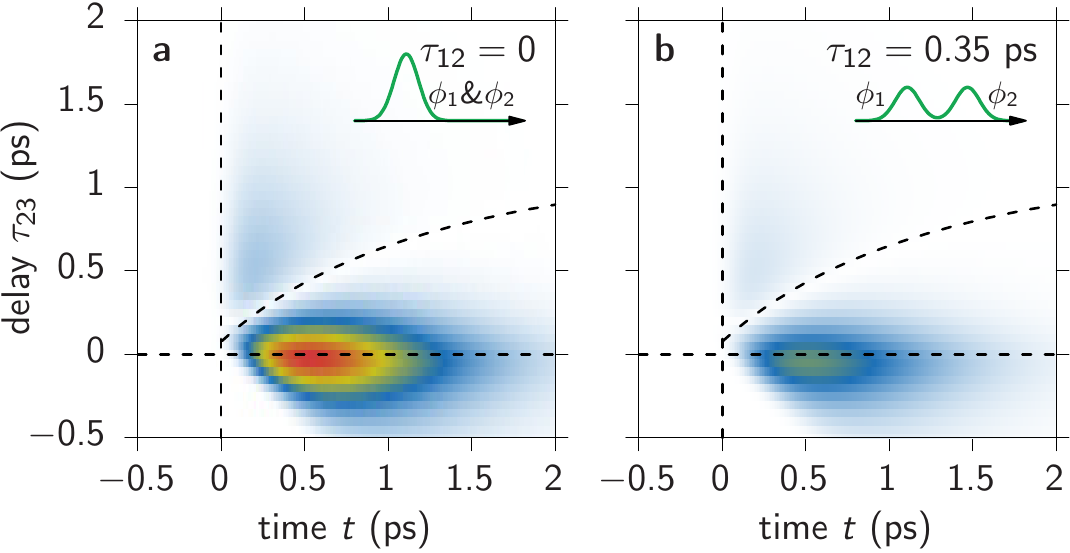}
	\caption{Impact of a non-vanishing pulse delay $\tau_{12}$ on the SWM amplitude dynamics. (a) Normalized SWM amplitude dynamics for $\tau_{12}=0$ from Fig.~4(b) in the main text. (b) SWM amplitude dynamics for $\tau_{12}=0.35$~ps normalized to the maximum value of (a). The insets schematically show the timing of the pulses with $\phi_1$ and $\phi_2$.}\label{fig:SWM_tau12}
\end{figure}

In the main text we have considered a vanishing delay between the two pulses labeled with $\phi_1$ and $\phi_2$. As a reference the corresponding SWM dynamics are shown in Fig.~\ref{fig:SWM_tau12}(a) as functions of $\tau_{23}$. To identify the impact of a non-vanishing delay between these two pulses, in Fig.~\ref{fig:SWM_tau12}(b) we plot the SWM dynamics for $\tau_{12}=0.35$~ps. As depicted in the inset, for this delay the two pulses are almost entirely separated. The corresponding SWM amplitude dynamics are identical to the ones with $\tau_{12}=0$ in Fig.~\ref{fig:SWM_tau12}(a) but have an overall smaller amplitude. This demonstrates that the destructive echo is not caused by pulse overlap effects. The reason for the decreased amplitude is, that the coherence $^{\phi_1}p_1^+$ after the first arriving pulse declines due to the considered dephasing. Consequently, the following processes, i.e., optical excitations and local-field mixing, are the same but start from a smaller coherence. Therefore, the final SWM signal is weaker.

\section{Impact of inhomogeneous broadening}\label{sec:echo}
\begin{figure}[h]
	\centering
	\includegraphics[width=\columnwidth]{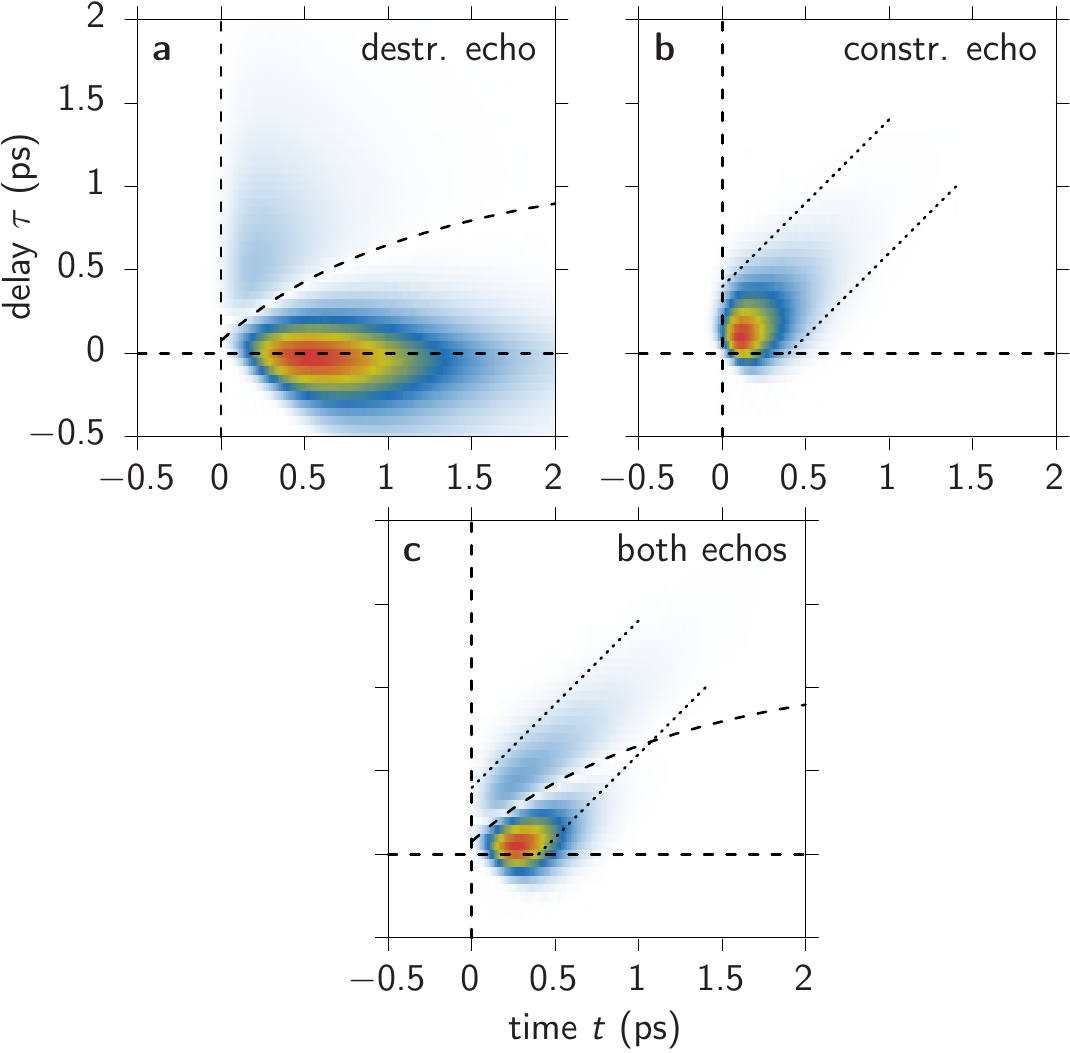}
	\caption{Impact of a non-vanishing inhomogeneous broadening on the SWM amplitude dynamics. (a) SWM amplitude dynamics with destructive echo from Fig.~4(b) in the main text. (b) SWM amplitude dynamics for $V=0$ but with an inhomogeneous broadening of $\sigma=4$~ps$^{-1}$ resulting in a constructive photon echo marked by the dotted black lines. (c) Combination of constructive and destructive photon echo. The signal depression is only visible in the vicinity of the constructive echo, while the entire signal is suppressed for all other times.}\label{fig:SWM_inhom}
\end{figure}

The experimental investigations of the sample do not show a significant inhomogeneous broadening, which should have manifested in the development of a photon echo in FWM and SWM. To demonstrate the impact of inhomogeneous broadening on the visibility of the destructive echo, we plot the destructive photon echo dynamics in Fig.~\ref{fig:SWM_inhom}(a), where the curved dashed black line indicates the minimum of the signal. In Fig.~\ref{fig:SWM_inhom}(b) we simulate the same SWM signal but set the local field coupling to $V=0$ and consider an inhomogeneous broadening of $\sigma=4$~ps$^{-1}$ which leads to the development of a traditional constructive photon echo~\cite{langbein2010coherent}. As marked by the two black dotted lines, the signal stretches along the diagonal $\tau=t$. Figure~\ref{fig:SWM_inhom}(c) combines both phenomena, showing the SWM including the original $V=100$~ps$^{-1}$ and the inhomogeneous broadening $\sigma=4$~ps$^{-1}$. As indicated by the dashed and dotted lines, the destructive photon echo is still clearly visible but only in the time span of the traditional constructive photon echo.

\section*{References}
